\documentclass[aps,pre,twocolumn,showpacs,superscriptaddress,nofootinbib]{revtex4-2}  
\usepackage{graphicx}  
\usepackage[caption=false,subrefformat=parens,labelformat=parens]{subfig}
\usepackage{float}
\usepackage[toc,page]{appendix}
\usepackage[table]{xcolor}
\usepackage{mathtools}
\usepackage{dcolumn}   
\usepackage{bm}        
\usepackage{color}

\usepackage{hyperref}
\usepackage{pifont}
\usepackage{mathrsfs}
\usepackage{amsmath, amsthm, amssymb}
\usepackage{amsfonts}
\usepackage{relsize}    
\usepackage{bbold}       
\usepackage{accents}    

\hypersetup{
	colorlinks=true, 
	linktoc=all,     
	linkcolor=blue,  
	citecolor=blue,
	filecolor=blue,
	urlcolor=blue
}
\usepackage{soul}
\newtheorem*{theorem}{Theorem}

\newcommand{\beq}{\begin{equation}}
	\newcommand{\eeq}{\end{equation}}
\newcommand{\bqa}{\begin{eqnarray}}
	\newcommand{\eqa}{\end{eqnarray}}

\newcommand{\erf}[1]{Eq.~(\ref{#1})}

\newcommand{\erfa}[2]{Eqs.~(\ref{#1}) and (\ref{#2})}
\newcommand{\arf}[1]{{App.}~\ref{#1}} 
\newcommand{\srf}[1]{Sec.~\ref{#1}} 
\newcommand{\crf}[1]{Ref.~\cite{#1}} 
\newcommand{\trf}[1]{Table ~\ref{#1}}

\newcommand{\frf}[1]{Fig.~\ref{#1}}

\newcommand{\ea}{{\it et al.}}

\newcommand{\dg}{^\dagger}

\definecolor{BLACK}{gray}{0}
\definecolor{RED}{rgb}{1,0,0}
\definecolor{GREEN}{rgb}{0.2,.6,0.2}
\definecolor{amber}{rgb}{1.0,0.49,0.0}

\renewcommand{\[}{\left[}

\newcommand{\an}[1]{\left\langle{#1}\right\rangle}

\newcommand{\smallfrac}[2]{\mbox{$\frac{#1}{#2}$}}

\newcommand{\bra}[1]{\langle{#1}|}

\newcommand{\ket}[1]{|{#1}\rangle}

\newcommand{\op}[2]{\ket{#1}\bra{#2}}

\begin{document}
	
	\widetext

	
	\title{Quantum approximate optimization of finite-state bosonic systems}

	\author{Shakib Daryanoosh} 
\email{shakib.daryanoosh@curtin.edu.au}
\affiliation{Curtin Centre for Optimisation and Decision Science, Curtin University, Whadjuk Country, Perth 6102, Australia}

%
%


%
\vskip 0.25cm


	\begin{abstract} 
		There exist numerous problems in nature inherently described by finite $D$-dimensional states. Formulating these problems for execution on qubit-based quantum hardware requires mapping the qudit Hilbert space to that of multiqubit which may be exponentially larger. To exclude the infeasible subspace, one common approach relies on penalizing the objective function. However, this strategy can be inefficient as the size of the illegitimate subspace grows. Here we propose to employ the Hamiltonian-based quantum approximate optimization algorithm (QAOA) through devising appropriate mixing Hamiltonians such that the infeasible configuration space is ruled out. We investigate this idea by employing binary, symmetric, and unary mapping techniques. It is shown that the standard mixing Hamiltonian (sum of the bit-flip operations) is the optimal option within the standard universal gate set for symmetric mapping, where the controlled-NOT gate count is used as a measure of implementation cost. In contrast, the other two encoding schemes witness a $p$-fold increase in this figure for a $p$-layer QAOA. We apply this framework to quantum approximate thermalization and find the ground state of the repulsive Bose-Hubbard model in the strong and weak interaction regimes.
	\end{abstract} 
	
	\maketitle

	\section{Introduction} \label{sec:intro}
	The quest for building utility-scale quantum computers has intensified in the past few years \cite{KimKan23,BraThe24,AleZho25}. Despite significant efforts, fault-tolerant quantum hardware might still be out of reach for the foreseeable future \cite{MohMar25}. In near- to mid-term, noisy quantum processors enable the implementation of small-scale algorithms without performing quantum error correction \cite{Pre18,BhaGuz22}. Leveraging these devices, variational quantum algorithms (VQAs) are promising as their shallow depth quantum circuits make them more resilient to noise in comparison to fault-tolerant architectures \cite{CerCol21}. 
	
	One of the leading candidates in this family of algorithms is the quantum approximation optimization algorithm introduced by Farhi \ea ~\cite{FarGut14}. This is a hybrid algorithm that combines quantum routines with classical optimization techniques to find approximate solutions for combinatorial optimization and eigenvalue problems. This heuristic algorithm has been applied to a large body of works that can be formulated in terms of binary variables. 
	
	There are a wide variety of interesting problems that are inherently finite dimensional and may naturally not fit in this framework. One prominent example is studying quantum mechanical systems arising in physics, chemistry, and material science. Achieving this goal on qubit-based quantum processors requires mapping from the original higher-dimensional qudit states onto multi-qubit ones. Various encoding schemes have been examined in this context \cite{SawGue20, MatNav21, PenRoe25}. However, the mapping is often not one-to-one, and consequently there exist an infeasible subspace that should be ruled out of calculations (for a $K$-qubit system with the state space of size $2^K$ only $D$ eigenstates are utilized). One approach to exclude such redundant states is based on adding a penalty term to the formalism (the objective function) \cite{RyaIzm19, LyuBay24}. However, for practical scenarios the size of the illegitimate subspace is exponentially larger than that of the physical. As a result, searching for the candidate solution becomes inefficient by exploring this huge configuration. 
	
	In this work we tackle this issue by confining the search to the feasible subspace in the context of QAOA setting. The crucial step is to carefully construct suitable parameterized quantum circuits that restrict the  simulations within the space of allowed states. In a $p$-layer QAOA, each round is consist of applying two unitary operations: one based on the problem (cost) Hamiltonian $\hat{H}_C$ and the other generated by some mixing (driving) Hamiltonian $\hat{H}_M$ (also know as \emph{mixer}). The latter is usually set to \erf{eqn:Hm:std}, as it was proposed in the original work \cite{FarGut14}. For problems we are concerned with in this paper, the key idea is to devise specific mixers that preserve the feasible subspace. 
	
	In fact, this procedure can be classified as the Hamiltonian-based QAOA from the extended version of the original QAOA, known as the quantum alternating operator ansatz (likewise acronymed as QAOA) \cite{HadBis19}. This method has already been applied to graph-coloring problems to satisfy the constraint conditions \cite{ZhiRie20} and quantum chemistry \cite{KreTub21}. Here we investigate quantum approximate optimization of $D$-dimensional bosonic systems using three different mappings, namely the binary, symmetric, and unary encoding schemes. Since entangling operations are costly to realize compared to single gates, we employ the controlled-not gate count as an implementation cost measure. 
	
	Overall, we find that symmetric mapping performs better than the other two. This is demonstrated by considering entangling resources needed for initial state preparation, implementing mixing Hamiltonian and the final measurement stage in the relevant basis. In particular, we observe the standard mixer, \erf{eqn:Hm:std}, is the best choice for symmetric-encoded problems. Given the $p$-fold increase in the number of required CNOT gates for the binary and unary mappings, and so long empirical factors matter, this outcome promotes the symmetric encoding to the top. Throughout this work, the term “optimal” is used exclusively in reference to the symmetric encoding, and denotes optimality with respect to entangling-gate requirements at the encoding and Hamiltonian level, within the assumed gate set and cost model. In particular, the standard mixer for the symmetric encoding factorizes into single-qubit Pauli-$X$ operations and therefore requires no entangling gates under any compilation consistent with these assumptions.
	
	More recently, it has been shown that VQAs can be used to generate certain many-body quantum states \cite{HoHsi19, WauSan20}. Here, we firstly inspect quantum approximate thermalization within the QAOA framework \cite{VerBia19,VerHid19,WuHsi19,ZhuMon20,MatPap21, DieGar23}. Quantum thermalization is important from both the fundamental and practical perspectives. For instance, it assists with understanding the emergence of statistical mechanics from closed quantum systems \cite{Ueda20}. This phenomenon also found applications in quantum machine learning for training generative adversarial network models \cite{AmiMel18, VerBia19}. As an example, we consider a system of coupled harmonic oscillators and show that how to approximately generate the target thermal (Gibbs) state at inverse temperature $\beta$. The key step involves initializing the protocol at the thermal state of the mixing Hamiltonian. In principle, the system should stay close to the instantaneous Gibbs state until it reaches the (near) thermal equilibrium.   
	
	Secondly, we use QAOA to prepare lowest-energy state of systems governed by the Bose-Hubbard model. This method has become the workhorse for discovering phenomena such as the superfluid-to-Mott-insulator phase transition \cite{KuhMon98,GreBlo02}. It provides a framework to scrutinize correlated bosonic systems, and its extension is capable of simulating complex quantum systems \cite{BloNas12}. We discuss the impact of truncating the Hilbert space on choosing appropriate parameter regimes. Our observation is that the proposed method converges to the system's ground state with far less resources (classical and quantum) in the strong interaction regime. In contrast, simulating the weak interaction domain needs a deeper and perhaps more expressive ansatz. This can be related to the highly quantum correlated structure of the ground state in the kinetically dominated regime. We note that these numerical examples are intended to illustrate general, encoding-dependent principles rather than to establish scaling performance claims; the qualitative conclusions regarding entangling-gate costs follow from the structure of the encodings and are not specific to the instances shown.
	
	The rest of the paper is organized as follows. In \srf{GQAOA} we briefly review the QAOA structure. Section \ref{sec:qudb} presents the encoding schemes. Section \ref{sec:Hmix} starts with outlining a class of mixing Hamiltonians for $D=3$ dimensional systems, then provides gates analysis prior to discussing $D>3$ scenarios.  Section \ref{sec:bosonic} applies the formalism to bosonic systems followed by investigating two case studies. We conclude by summarizing the findings in \srf{sec:conclusion}.

	\section{From QAOA to QAOA} \label{GQAOA}
	For completeness, we briefly summarize the standard QAOA framework, focusing only on elements needed to introduce the encoding- and mixer-dependent constructions studied in this work. The quantum approximate optimization algorithm was originally proposed to solve combinatorial optimization problems \cite{FarGut14}. It is a hybrid quantum-classical variational algorithm, where the problem is mapped to a cost Hamiltonian $\hat{H}_C$ whose ground state encodes the solution. Evolution under $\hat{H}_C$ alone leaves $\langle \hat{H}_C \rangle$ invariant, so a second noncommuting Hamiltonian $\hat{H}_M$, known as the mixer, is introduced to explore the solution space.
	
	The starting point involves preparation of the ground state of the standard driving Hamiltonian 
	\begin{equation} \label{eqn:Hm:std}
		\hat{H}_M = \sum_{k=0}^{K-1} X_k
	\end{equation}
	where ${X}_k \coloneq I_{K-1}\otimes \cdots \otimes I_{k+1}\otimes X_k\otimes I_{k-1}\otimes\cdots\otimes I_0$ is the Pauli $X$ operator acting on the $k-$th qubit. Here we use little endian convention where lower indices represent the least significance.
	Through alternate applications of the unitaries $\hat{U}_C(\gamma) = {\rm Exp}[-i \gamma \hat{H}_C]$ and $\hat{U}_M(\nu) = {\rm Exp}[{-i \nu \hat{H}_M}]$ to the initial state $\ket{\Psi_{\rm in}}$ a trial state can be constructed:
	\begin{equation}
		\ket{\Psi (\Theta)} = \hat{U}_M(\nu_p) \hat{U}_C(\gamma_p) \cdots \hat{U}_M(\nu_1) \hat{U}_C(\gamma_1) \ket{\Psi_{\rm in}},
	\end{equation} 
	where $\Theta\equiv(\boldsymbol{\gamma,\nu}) = (\gamma_p,\cdots,\gamma_1,\nu_p,\cdots, \nu_1)$ is the set of $2p$ real parameters for $p$ layers. The classical optimization problem is to find $\Theta^*$ minimizing $f(\Theta) \coloneq \langle \ket{\Psi (\Theta)} |\hat{H}_C | \ket{\Psi (\Theta)} \rangle$. This is accomplished via measurements on $\ket{\Psi(\Theta)}$ and classical optimization. Whether this solution is (near) optimal, critically depends on the discovery of appropriate values for $\Theta$ leading to approximate ground state. We note that strategies for choosing good initial parameters are under active research \cite{McCNev18,LeeAsa24,HaoPis25}.
	
	Standard practice uses the simple initial state and the mixer in Eq.~\eqref{eqn:Hm:std}. Constraints are often incorporated as penalty terms in the cost Hamiltonian, but this enlarges the search space. Following approaches in quantum adiabatic computation \cite{HenSpe16,HenSar16}, and later in the extended version of QAOA  know as quantum alternating operator ansatz \cite{HadBis19}, one can instead design mixers that preserve the feasible subspace. 	
	
	The extended QAOA allows more general mixing operations, which may not correspond to unitary time evolution under a Hamiltonian. A subclass of this family still uses mixers of the form ${\rm Exp}[{-i \nu \hat{H}_M}]$; our work focuses exclusively on such mixers. In the following sections, we exploit this framework to compare how different encodings of finite-state bosonic systems affect the structure and resource cost of constraint-preserving mixers.

	\section{Qudit to multi-qubit mapping} \label{sec:qudb}
	Consider a collection of $L$ identical systems each of dimension $D$, such that the set $\{\ket{\Psi_d^{\ell}}\}$ for $0\le d\le D-1$ represents an orthonormal basis for the $\ell$-th system where $1\le \ell\le L$. For the purpose of reviewing different mapping methods we set $L=1$ without loss of generality. 
	
	\subsubsection{Binary encoding}
	The standard method for encoding a qudit into $K= \lceil\log_2 D\rceil$ qubits is the hardware efficient binary mapping. Here $\lceil . \rceil$ denotes the ceiling function. For scenarios in which $D=2^K$, states from the qudit and qubits spaces are fully mapped onto each other. In general, 
	
	\begin{equation} \label{ket:map:b}
		\ket{\Psi_d} \longmapsto \ket{\psi_d^K}_{b} \equiv \ket{b_{K-1} b_{K-2} \cdots b_1 b_0},
	\end{equation}
	where $b_k \in \{0,1\}$ are binary variables, and 
	\begin{equation}
		d = \sum_{k=0}^{K-1} 2^k b_k.
	\end{equation}
	Therefore, a mapping onto the qubits space can be constructed according to
	\begin{equation} \label{bin:map:op} 
		\hat{\cal{M}}_b = \sum_{d=0}^{D-1} \ket{\psi_d^K}_b \bra{\Psi_d}.
	\end{equation} 
	It is important to note that if the transformation is not one-to-one ($D<2^K$), then an infeasible subspace exists such that 
	\begin{equation} \label{not:res:id}
		\hat{\cal{M}}_b\dg \hat{\cal{M}}_b = I_D \neq \hat{\cal{M}}_b \hat{\cal{M}}_b\dg.
	\end{equation} 
	Operators between the two spaces can be converted via 
	\begin{equation} \label{op:map}
		\hat O_b = \hat{\cal{M}}_b \hat O \hat{\cal{M}}_b\dg.
	\end{equation}
	We will see in \srf{sec:bosonic} the relevance of binary encoding for Fock states.  
	
	\subsubsection{Symmetric encoding}
	In physics the existence of symmetry indicates conservation of some physical quantity according to Noether's theorem \cite{Noe71}. Even though different mapping methods retain this property, the symmetric encoding should be able to better highlight the characteristics of conserved quantities. In symmetrical mapping, each basis state $\ket{\Psi_d}$ in the original space is mapped onto a state with Hamming weight $d$. In other words, basis elements are superposition of states with exactly $d$ qubits in $\ket{1}$: 
	\begin{equation} \label{ket:map:s}
		\ket{\Psi_d} \longmapsto \ket{\psi_d^K}_{s} \equiv \frac{1}{\sqrt{\binom{K}{d}}} \sum_{\substack{\rm Hamming \\ {\rm weight} \;d}} \ket{b_{K-1} \cdots b_1 b_0},
	\end{equation}
	where $K=D-1$, and the summation is over all binary kets with Hamming distance $d$. Transformation between the bases can be carried out similar to \erf{bin:map:op}, and the condition in \erf{not:res:id} always holds since there is no one-to-one mapping unless $D=2$. Operator can be represented in the symmetric basis via \erf{op:map} with the corresponding map.

	\subsubsection{Unary encoding}
	For symmetric mapping we noticed the notion of resource efficiency is no longer valid as the number of qubits required for encoding grows exponentially with $D$. This redundancy might seem unnecessary, but it can be essential to protect quantum information by correcting errors due to noisy hardware. Another mapping scheme with similar excessiveness is the unary encoding onto a register of $K=D$ qubits:
	\begin{equation} \label{ket:map:u}
		\ket{\Psi_d} \longmapsto \ket{\psi_d^K}_{u} \equiv \ket{0_{K-1}\cdots 0_{d+1} 1_{d} 0_{d-1}\cdots 0_1 0_0 }.
	\end{equation}
	In this representation only one qubit is in the logical state $\ket{1_d}$ and its index refers to that of the original space's state. This feature becomes useful in \srf{sec:bosonic} where the label serves as the number of bosons in a bosonic mode.
	
	\begin{table*}
		\caption{\label{tbl:cnot:d3} Summary of CNOT gate counts for compilation of $e^{-i \nu \hat{H}_{M,m}^{(j)}}$ for different encoding schemes. For a $p$-layer QAOA this figure is $p$-fold higher. The amount of entanglement in the initial state is quantified by the logarithmic negativity, \erf{eqn:LN:gen}. Consult \srf{sec:Hmix:multiq} for the discussion related to performing entangling measurements. Here we set $D=3$. }
		\begin{tabular}{ c c c  c c c  c c c c  c c c}
			\hline \hline 
			\multicolumn{3}{c}{\rm Encoding}\quad & \multicolumn{3}{c}{\rm Binary} &  \multicolumn{4}{c}{\qquad \quad \rm Symmetric} &  \multicolumn{3}{c}{\qquad \quad\rm Unary} \\ \hline
			\multicolumn{3}{c}{\begin{tabular}{@{}c@{}} {\rm Mixing} \\ {\rm Hamiltonian} \end{tabular}}\quad &\; $\hat{H}_{M,b}^{(1)}$\;\;& \;$\hat{H}_{M,b}^{(2)}$\;\; &\; $\hat{H}_{M,b}^{(3)}$\;\;&\qquad  $\hat{H}_{M,s}^{(1)}$\;\;&\; $\hat{H}_{M,s}^{(2)}$\;\; & \;$\hat{H}_{M,s}^{(3)}$\;\; &\; $\hat{H}_{M,s}^{\rm Opt}$\;\;&\qquad  $\hat{H}_{M,u}^{(1)}$\;\;& \;$\hat{H}_{M,u}^{(2)}$ \;\;& \;$\hat{H}_{M,u}^{(3)}$ \\[14pt] 
			\multicolumn{3}{c}{\# {\rm CNOT} ($\times p$)}\quad&\;2\;\;&\;2\;\;&\;4\;\;&\qquad  4\;\;&\;4\;\;&\;4\;\;&\;0\;\;&\qquad  12\;\;&\;8+4\;\;&\;12\\[6pt]
			\multicolumn{3}{c}{$LN (\hat\rho_{\rm in})$}\quad &\;0\;\;&\;0\;\;&\;1\;\;&\qquad 0.58\;\; &\; 1\;\; & \;0.58 \;\;& \;0\;\; &\qquad {\begin{tabular}{@{}c@{}} $LN_{2|4}=0$ \\ $LN_{4|2}=1$ \end{tabular}}\;\; & \;{\begin{tabular}{@{}c@{}} $LN_{2|4}=1$ \\ $LN_{4|2}=1$ \end{tabular}}\;\; & 
			\;{\begin{tabular}{@{}c@{}} $LN_{2|4}=1$ \\  $LN_{4|2}=0$ \end{tabular}}\;\;\\
			\multicolumn{3}{c}{\begin{tabular}{@{}c@{}} {\rm Entangling} \\ {\rm measurement} \end{tabular}}\quad &\;\ding{55}  \;\;&\;\ding{55} \;\;&\; \ding{55} \;\;&\qquad \ding{51} \;\;&\;\ding{51} \;\;&\;\ding{51} \;\;&\;\ding{51} \;\;&\qquad \ding{55} \;\;&\;\ding{55} \;\;&\;\ding{55} \\
			\hline \hline
		\end{tabular}
	\end{table*}
	
	\section{Mixing Hamiltonian} \label{sec:Hmix}
	To restrict the system evolution to the allowed subspace, the generator of  dynamics should keep the system out of the infeasible subspace. Therefore, the mixing Hamiltonian can be formalized by the linear combination of (\emph{partial}-mixing) operators constructed by basis states of the multi-qubit subspace:
	\begin{equation} \label{mixer:gen}
		\hat{H}_{M,m} = \sum_{d<d'}^{D-1} \ket{\psi_d^K}_m\bra{\psi_{d'}^K} + \textrm{h.c.} \equiv \sum_{d<d'}^{D-1} \hat{H}_{M,m}^{(d+d')}. 
	\end{equation}
	In what follows we consider the simplest case of $D=3$ level systems. We will see that various driving Hamiltonian may be selected depending on the encoding scheme.

	\subsection{Three-state systems}
	\emph{Binary.---}
	The basis states from the original to multi-qubit spaces are mapped based on \erf{ket:map:b}:
	\begin{equation} \label{ket:map:b:d3}
		\ket{\Psi_0} \mapsto \ket{00}, \quad \ket{\Psi_1} \mapsto \ket{01}, \quad \ket{\Psi_2} \mapsto \ket{10}. 
	\end{equation} 
	In this example, the infeasible subspace is small compared to the dynamical space. Nevertheless, to avoid the former, one can drive the system to the feasible subspace through using a class of mixing Hamiltonians given below: 
	\begin{subequations}
		\begin{eqnarray}
			\hat{H}_{M,b}^{(1)} &=& \frac{1}{2}(IX+ZX), \label{H:mix:d3s1} \\
			\hat{H}_{M,b}^{(2)} &=& \frac{1}{2}(XI+XZ), \label{H:mix:d3s2}\\
			\hat{H}_{M,b}^{(3)} &=& \frac{1}{2}(XX+YY). \label{H:mix:d3s3}
		\end{eqnarray}
	\end{subequations}
	Here the 2-qubit operation $ZX$ generates rotation about $ZX$ axis by angle $\nu$ which can be interpreted as a uniformly controlled $\hat{R}_x$ gate:
	\begin{equation}
		\hat{R}_{ZX}(\nu) = e^{-\frac{i}{2}\nu ZX} =
		\begin{pmatrix}
			\hat{R}_{X}(\nu)&0 \\
			0&\hat{R}_{X}(-\nu)
		\end{pmatrix},  
	\end{equation} 
	where we used block diagonal representation, and $\hat{R}_X(\nu) = e^{-i\nu X/2} = {\rm cos}(\nu/2) I -i {\rm sin}(\nu/2) X$. Therefore, the unitary operation generated by the first mixing binary Hamiltonian in the computational basis $\ket{b_1 b_0}$ is
	\begin{equation}
		e^{-i\nu \hat{H}_{M,b}^{(1)}} =
		\begin{pmatrix}
			{\rm cos}(\nu)& -i {\rm sin}(\nu)&0&0 \\
			-i {\rm sin}(\nu)& {\rm cos}(\nu)&0&0 \\
			0&0&1&0 \\
			0&0&0&1
		\end{pmatrix}.  
	\end{equation} 
	This transformation only mixes $\ket{00}$ and $\ket{01}$ and leaves the subspace spanned by $\ket{10}$ and $\ket{11}$ unchanged.
	
	One way to figure out whether to employ either of the partial-mixing Hamiltonians or a combination of them $\sum_{j\in A}  \hat{H}_{M,b}^{(j)}$ where $A = \{1,\cdots,D(D-1)/2\}$, is through the logical gate complexity analysis. This can be achieved by decomposing the unitary associated with the bias Hamiltonian to single- and two-qubit operations. The entangling CNOT gates are costly due to higher error rates and longer execution times in comparison to single-qubit gates. Therefore, we use the CNOT gate count for the purpose of optimizing the ansatz quantum circuit. For this analysis the IBM's Qiskit with the first-order Lie-Trotter approximation is employed \cite{JavGam24}. It turns out the $XY$ mixer, \erf{H:mix:d3s3}, requires double the number of CNOT gates compared to the other two choices, \erfa{H:mix:d3s1}{H:mix:d3s2}, consult \trf{tbl:cnot:d3}.

	\emph{Symmetric.---}
	According to \erf{ket:map:s}, the mapping between the 3-state and two-qubit basis states is the following:
	
	\begin{equation} \label{ket:map:s:d3}
		\ket{\Psi_0} \mapsto \ket{00}, \quad \ket{\Psi_1} \mapsto \frac{1}{\sqrt{2}}\left(\ket{01} + \ket{10} \right), \quad \ket{\Psi_2} \mapsto \ket{11}. 
	\end{equation} 
	Similar to the binary encoding, the infeasible subspace is not large, and it only consists of the antisymmetric state $\frac{1}{\sqrt{2}}\left(\ket{01} - \ket{10} \right)$. The mixing Hamiltonians are:
	\begin{subequations}
		\begin{eqnarray}
			\hat{H}_{M,s}^{(1)} &=& \frac{1}{2\sqrt{2}}(IX+XI+ZX+XZ),\\
			\hat{H}_{M,s}^{(2)} &=& \frac{1}{2}(XX-YY),\\
			\hat{H}_{M,s}^{(3)} &=& \frac{1}{2\sqrt{2}}(IX+XI-ZX-XZ).
		\end{eqnarray}
	\end{subequations}
	Note that the linear combination 
	\begin{equation} \label{H:mix:sym:opt:d3}
		\hat{H}_{M,s}^{\rm Opt} \equiv \hat{H}_{M,s}^{(1)}+\hat{H}_{M,s}^{(3)} = \frac{1}{\sqrt{2}} \sum_{j=0}^{1} X_j,
	\end{equation}
	reproduces the standard mixing Hamiltonian usually employed in the original QAOA, and $\hat{H}_{M,s}^{(2)}$ is equivalent to the XY-mixer introduced in the extended QAOA. Here, each unitary ${\rm Exp}[{-i\nu \hat{H}_{M,s}^{(j)}}]$ Trotterization requires 4 entangling CNOT gates. However, the standard mixer $\sum_j X_j$ is an optimal option (within the standard universal gate set and cost model) that rules out the infeasible antisymmetric state $\ket{\psi_3}_s = \frac{1}{\sqrt{2}}(\ket{01}-\ket{10})$ according to the following unitary evolution (expressed in the symmetric basis $\{\ket{\psi_0}_s, \ket{\psi_1}_s,\ket{\psi_2}_s, \ket{\psi_3}_s\}$):
	\begin{equation}
		e^{-i\nu \sqrt{2} \hat{H}_{M,s}^{\rm Opt}} =
		\begin{pmatrix}
			{\rm cos}^2(\nu)& \frac{-i}{\sqrt{2}} {\rm sin}(2\nu)& -{\rm sin}^2(\nu)&0 \\
			\frac{-i}{\sqrt{2}} {\rm sin}(2\nu)& {\rm cos}(2\nu)& \frac{-i}{\sqrt{2}} {\rm sin}(2\nu)&0 \\
			-{\rm sin}^2(\nu)& \frac{-i}{\sqrt{2}} {\rm sin}(2\nu)& {\rm cos}^2(\nu)&0 \\
			0&0&0&1
		\end{pmatrix},  
	\end{equation} 
	and only needs single-qubit rotations $\hat{R}_{X_j}$. We will see below this feature holds irrespective of the value $D$. This is appealing from the viewpoint of experimental realization, because for high $D$ the binary and unary encoding schemes demand exponentially more CNOT gates, see \frf{fig:CNOT:Hm}. This is one of the main results of this work. 
	
	\begin{figure}
		\includegraphics[width=\linewidth]{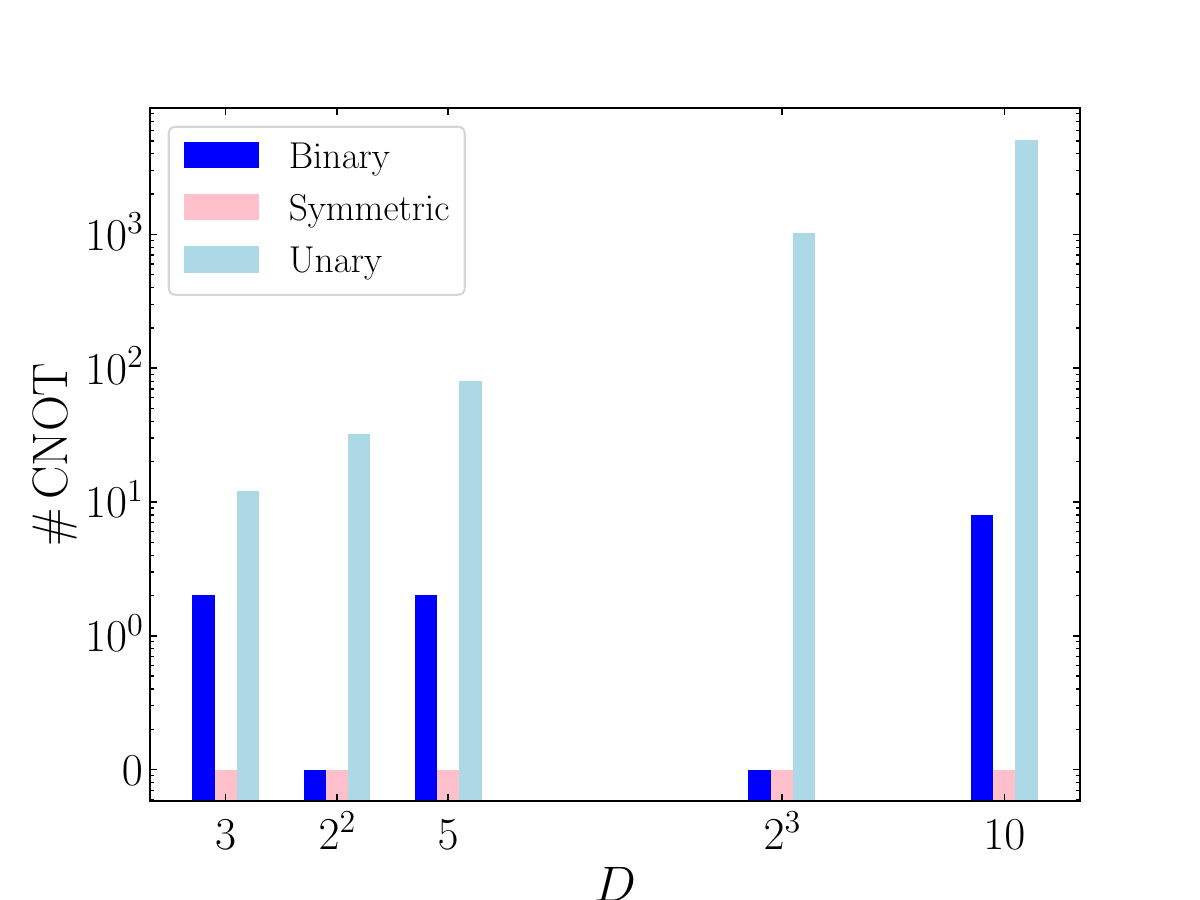}
		\caption{\label{fig:CNOT:Hm} (Color online). The number of CNOT gates required by a best candidate mixing Hamiltonian for $D$-state systems for binary (dark blue), symmetric (pink) and unary (light blue) encoding schemes. Note that the symmetric mapping is overall more efficient than the other two encoding techniques. For the sake of clarity, zero count of the entangling gate is set to a small number ($0.1$).
		}
	\end{figure}
	
	\emph{Unary.---}
	As expected with this encoding method, the Hilbert space is larger compared to the other two approaches. In this case the basis states are mapped via:
	\begin{equation}
		\ket{\Psi_0} \mapsto \ket{001}, \quad \ket{\Psi_1} \mapsto\ket{010}, \quad \ket{\Psi_2} \mapsto \ket{100},
	\end{equation} 
	and the mixing Hamiltonians are 
	\begin{subequations}
		\begin{eqnarray}
			\hat{H}_{M,u}^{(1)} &=& \frac{1}{4}(IXX+IYY+ZXX+ZYY), \label{H:una:d3s1}\\
			\hat{H}_{M,u}^{(2)} &=& \frac{1}{4}(XIX+YIY+XZX+YZY), \label{H:una:d3s2}\\
			\hat{H}_{M,u}^{(3)} &=& \frac{1}{4}(XXI+YYI+XXZ+YYZ). \label{H:una:d3s3}
		\end{eqnarray}
	\end{subequations}

	In contrast to the other encoding strategies, there is no term with only a single Pauli operator. In fact, the mapping structure necessitates three times as many CX gates for Trotterization of ${\rm Exp}[{-i\nu \hat{H}_{M,u}^{(j)}}]$. Although we use the CNOT count as a simple comparative measure across encodings, note that the unitary corresponding to \erf{H:una:d3s2} requires 4 of these gates between the zeroth and second qubits, whereas the other two only need nearest-neighbor qubit coupling. For high-dimensional systems, realizing the required long-range interaction can be empirically challenging either due to the physical layout or noise level. Other than qubit connectivity, there are factors such as native gate set and multiqubit operations, synthesis strategy, and initial qubit placement (network topology) that can affect this metric. 
	
	Table \ref{tbl:cnot:d3} summarizes the number of CNOT gates required for each mapping and corresponding mixing Hamiltonians. It is worth noting that there is a $p$-fold increase in the number of required CNOT gates for implementing $p$ rounds of QAOA. For the simple scenario we analyzed above, one might argue that due to the small size of the illegitimate subspace it makes sense to employ the standard mixing Hamiltonian $\sum_j X_j$ and penalize the cost function to avoid that subspace. For the symmetric encoding we already know this strategy is effective without resorting to penalizing the cost function (evolution is within the solution subspace). However, for this mapping the Bell measurements are needed after applying $p$ layers of the QAOA to calculate expectation values and/or to obtain measurement statistics. For unary encoding the idea of searching the full Hilbert space may reduce the computational cost even more given that the proposed mixing Hamiltonians require several times more CNOT gates (this figure is an order of magnitude when compared to the optimal symmetric case).

	\subsection{Many-state systems ($D>3$)} \label{sec:Hmix:multiq}
	For higher dimensional problems the trade off between penalizing the cost function and restricting dynamics to the solution subspace would favor the latter. There are three factors one should consider for choosing an encoding method: qubit resource, quantum error correction, and characteristics of the problem under study.
	\emph{Firstly}, hardware-efficient encoding techniques such as the binary mapping are desirable as they demand less qubits and therefore the infeasible subspace's dimension is exponentially smaller. \emph{Secondly}, quantum information is fragile and sensitive to noise and decoherence mechanism. It is widely believed that fault-tolerant quantum computation is essential for tackling large-scale practical problems. Therefore, detecting and correcting errors are key requirements for executing accurate quantum computing. This task demands many  physical qubits to collectively act like a logical qubit. The available large Hilbert space offered by the symmetric and unary encoding schemes provide an opportunity for error correction or mitigation. \emph{Lastly}, there are problems that possess inherent properties unique to them. For example, the quantum harmonic oscillator can be described by number states defined on the Fock space. These states can be appropriately mapped onto the binary encoding since each bitstring represents the number of field excitations (e.g., bosons). Alternatively, one could consider the crucial role of symmetry in physics leading to conservation laws via the Noether's theorem~\cite{Noe71}. Preserving certain system's characteristics through symmetric encoding can be more effective \cite{GarBar20, LyuBay24}.
	
	In light of the above considerations, the process of selecting an encoding procedure can be rather straightforward. Figure~\ref{fig:CNOT:Hm} shows the variation of CNOT operation count for a few different system dimensions $D$. The binary mapping shown in blue bars demonstrates a steady increase in the required number of CNOT gates except for  $D = 2^K$. For the latter, the mapping is one-to-one between the original and multi-qubit bases, meaning that there is not an infeasible subspace.  Therefore, the standard mixing Hamiltonian $\sum_{k=0}^{K-1} X_k$ may be chosen as the best option, see \frf{fig:CNOT:Hm} (for demonstration purposes, a small nonzero number $0.1$ is assigned to represent zero CNOT gate count). For each dimension a best candidate Hamiltonian with the lowest number of required CNOT gates is selected (in contrast to the three-state case where we list all possible nominated bias Hamiltonians in \trf{tbl:cnot:d3}). We note that the nominated mixer might not be optimal or unique with respect to the number of CNOT gates, unless for cases in which the Hilbert space is equidimensional for both the $D$-level and multi-qubit states (so long the entangling gate count is concerned). 
	
	\begin{figure}
		\includegraphics[width=\linewidth]{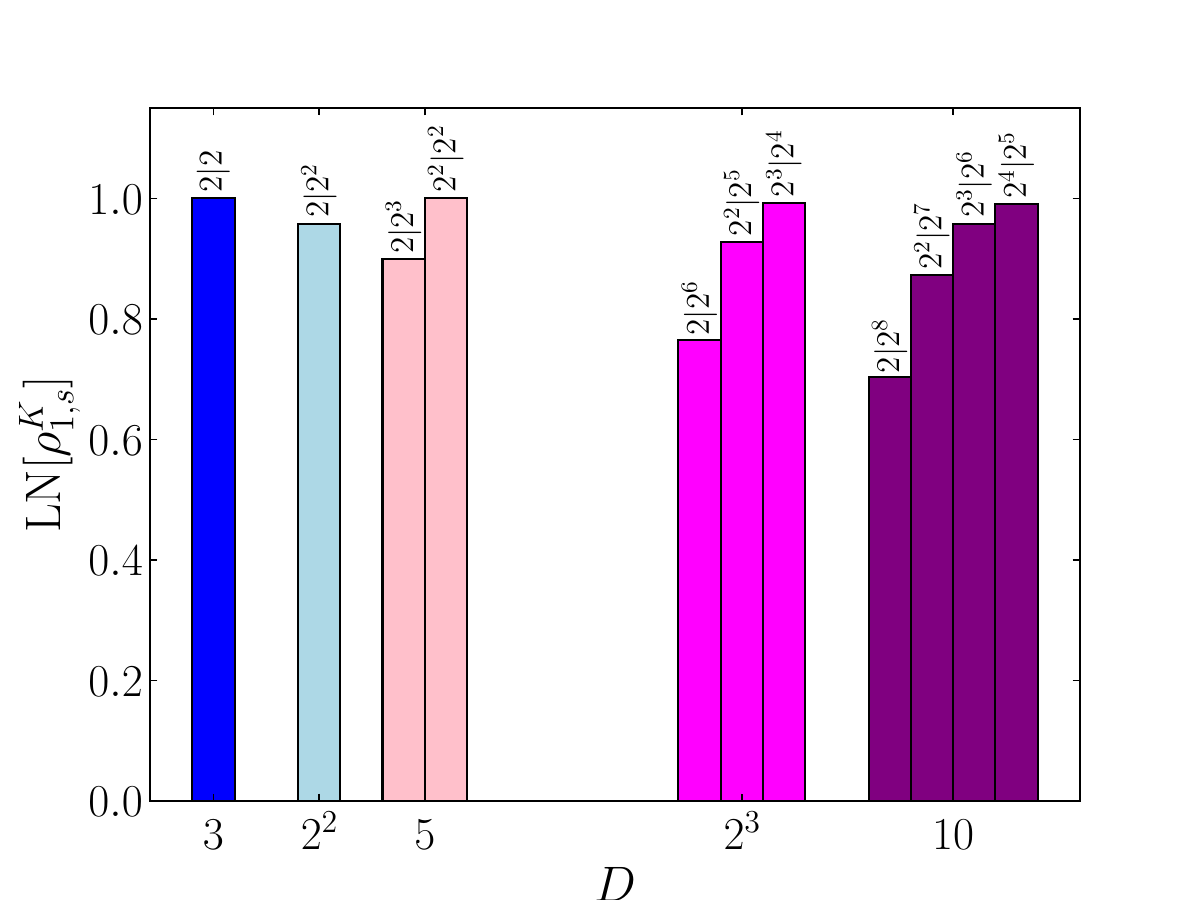}
		\caption{\label{fig:ln} (Color online). Logarithmic negativity of $\rho_{1,s}^K = \ket{\psi_1^K}_s\bra{\psi_1^K}$, \erf{ket:map:s}, as a function of dimension $D=K+1$. The partial transpose is calculated with respect to the first $j$ qubits with Hilbert space bipartitioning $2^j|2^{K-j}$. 
		}
	\end{figure}
	
For symmetric encoding the standard mixer is optimal as it does not need any two-qubit entangling gate. The pink bars in \frf{fig:CNOT:Hm} depict this for various $D$ (here we choose again a small number for the purpose of illustration). The reason is that Hamiltonian $\hat{H}_{M,s}^{\rm Opt}$ does not change the relative phase of the computational basis states in $\{\ket{\psi_d^K}_s\}_{d=1}^{D-2}$, and the mapping utilizes the entire computational eigenvectors. In comparison to the binary and unary encoding schemes for which the CNOT gate count grows exponentially with the system dimension, the symmetric mapping becomes appealing. In what follows we show that the unitary dynamics generated by the collective $X$ Hamiltonian, \erf{eqn:Hm:std}, preserves the symmetric subspace. 
	
	\begin{theorem}[Invariance of the symmetric subspace under the standard mixer] 
		Let $\hat U_M(\nu) = e^{-i \nu \hat H_M}$, where $\hat H_M$ is defined according to \erf{eqn:Hm:std}. Then, this Hamiltonian and its associated unitary operation preserve the symmetric subspace
		\begin{equation}
			\mathscr{H}_{\rm sym} = \{ \ket{\psi} : \hat \wp_\ell \ket{\psi} = \ket{\psi}, \forall \ell \in S_K\}.
		\end{equation}
		where $\hat \wp_\ell$ is the permutation operator, and $S_K$ is the symmetry group of $K$ qubits. 
	\end{theorem}
	
	\begin{proof} For any permutation $\ell \in S_K$ 
		\begin{equation} \label{eq:perm:mixer}
			\hat \wp_\ell \hat H_M \hat \wp_\ell\dg = \sum_{k=0}^{K-1} \hat \wp_\ell X_k \hat \wp_\ell\dg = \sum_{k=0}^{K-1} X_{\ell(k)},
		\end{equation}
		where $X_{\ell(k)}$ denotes the $k-$th Pauli $X$ operator undergoing permutation $\ell$ \footnote{ The last equality in \erf{eq:perm:mixer} can be proven by acting $\hat \wp_\ell\dg$ on the computational basis state $\ket{b_{K-1} b_{K-2} \cdots b_1 b_0}$ followed by the application of $X_k$ and $\hat \wp_\ell$. The resulting total action is that of $X_{\ell(k)}$ directly applied to the same basis.}. Since the latter is a permutation of $\{0,1,\cdots,K-1\}$, we obtain $\sum_{k=0}^{K-1} X_{\ell(k)} = \sum_{k=0}^{K-1} X_{k} = \hat H_M.$ Therefore, the mixer commutes with all permutations\footnote{ Note that according to the Schur-Weyl duality, an operator commuting with all permutations is block diagonal with respect to the decomposition into irreducible $S_K$-representations. This property may be used to argue that an operator commuting with all permutations cannot map between inequivalent irreducible representations carried by infeasible subspaces, leading to \erf{eq:decouple:subspa}. We refer the interested reader to \crf{FulHar91}.}: $[\hat H_M, \hat \wp_\ell] = 0, \; \forall \ell \in S_K$. As a result, $\hat H_M \ket{\psi}$ is invariant under all permutations. In other words, the symmetric subspace remains invariant by the action of mixing Hamiltonian: $\hat H_M \mathscr{H}_{\rm sym} \subseteq \mathscr{H}_{\rm sym}$.
		Now, since $\hat U_M$ is an exponential function of $\hat H_M$, and exponentials obey the commutation relations, then $[\hat U_M, \hat \wp_\ell]=0, \; \forall \ell$. Consequently, the unitary dynamics also preserve the symmetric subspace, that is, $ \hat U_M \mathscr{H}_{\rm sym} \subseteq \mathscr{H}_{\rm sym}$. 
	\end{proof}

	It is now straightforward to show that $\hat H_M$ decouples the symmetric subspace from the infeasible subspace $\mathscr{H}_{\rm inf}$. Let us define the following symmetrization operator
	\begin{equation}
		\hat \Pi \coloneq \frac{1}{K!} \sum_{\ell \in S_K} \hat \wp_\ell, 
	\end{equation}
	with properties 1) $\hat \Pi^2 = \hat \Pi$, 2) $\hat \Pi \ket{\psi} = \ket{\psi}$ for any symmetric state, and 3) $\Pi \ket{\chi} = 0$ for any state $\ket{\chi} \in \mathscr{H}_{\rm inf}$ orthogonal to the feasible subspace. Employing the last two conditions together with the commutation relation $[\hat H_M, \hat\Pi] = 0$ leads to
	\begin{eqnarray} \label{eq:decouple:subspa}
		\bra{\chi}\hat H_M \ket{\psi} = 0,
	\end{eqnarray}
	indicating that the leakage out of the symmetric subspace is impossible under $\hat H_M$ and/or $\hat U_M$.

	\subsubsection*{Initial state and measurement discussion}
	To draw a fair comparison we take into account the initial state preparation and measurement steps. Ideally, the input state should be simple to generate. This is fortunately the case for the standard driving Hamiltonian, \erf{eqn:Hm:std}, with its ground state equal to the coherent superposition of computational basis states $\ket{-}^{\otimes K}$ where $\ket{-}$ is the eigenstate of $X$ Pauli operator. This makes the symmetric mapping favorable over the other two (only $K$ Hadamard gates are applied to the register of input qubits). Now the question is can all other initial states be produced without entangling operations? To quantify entanglement we employ the logarithmic negativity metric defined according to \cite{Ple05}:
	\begin{equation} \label{eqn:LN:gen}
		LN (\hat\rho) := {\rm log}_2 \Vert \hat{\rho}^{\rm PT} \Vert_1 ,
	\end{equation}
	where $\Vert \hat{O} \Vert_1 = {\rm Tr}[\sqrt{\hat{O}\dg \hat{O}}\,]$ denotes the trace norm (the sum of singular values of $\hat O$), and $\hat{O}^{\rm PT}$ is the partial transpose of $\hat{O}$ with respect to a subsystem. Table \ref{tbl:cnot:d3} lists the amount of quantum correlation for different initial states. It is easy to check that the ground state of $\hat{H}_{M,b}^{(3)}$ and $\hat{H}_{M,s}^{(2)}$ are the maximally entangled Bell states $(\ket{01}-\ket{10})/\sqrt{2}$ and $ (\ket{00}-\ket{11})/\sqrt{2}$, respectively. Calculating the logarithmic negativity measure for the unary-encoded mixers requires partitioning of the Hilbert space (see below explanation). In any case, nonzero logarithmic negativity indicates the existence of nonclassical correlation in the ground state.

	Measurements in the computational basis are taken to be ``free'' for the binary and unary encoding techniques. In marked contrast, entangling measurements are necessary for the symmetric encoding scenario.  Figure \ref{fig:ln} demonstrates the logarithmic negativity for states $\rho_{1,s}^K = \ket{\psi_1^K}_s\bra{\psi_1^K}$ with hamming distance $d=1$, \erf{ket:map:s}, for several problem sizes $D=K+1$. It is evident that depending on partitioning of the multi-qubit Hilbert space $\mathscr{H}^{\otimes K}$ into $\mathscr{H}^{\otimes j}|\mathscr{H}^{\otimes K-j}$, the amount of quantum correlation can be different. Here $\mathscr{H}^{\otimes j}$ denotes the tensor product subspace for the first $j$ qubits. To get an estimate of the number of CNOT gate required for implementing projective measurements onto the symmetric basis, we benchmark against $K$-qubit Greenberger–Horne–Zeilinger (GHZ) states $\ket{{\rm GHZ}} = \smallfrac{1}{\sqrt{2}} (\ket{0}^{\otimes K} + \ket{1}^{\otimes K})$ which can be generated by means of $K-1$ nearest neighbor CNOT operations \cite{GHZ89}. With this in mind, let us assume a quantum circuit with approximately $D-2$ entangling CNOT gates can realize symmetric basis measurements. Now, consider an example where $D=5$ and QAOA has $p$ repetitions. Then, we only count the number of required CNOTs for generating the initial state, executing the unitary generated by the mixing Hamiltonian, and implementing the final measurement. For the binary and unary mapping procedures this number is almost $2p$ and $80p$, respectively, as opposed to $3$ for the symmetric case. 
	
	Although our focus here was solely on mixing Hamiltonians, it is important to highlight the role of the cost Hamiltonian in preserving the feasible subspace. The structure of $\hat H_C$ is block-diagonal with respect to the feasible/infeasible Hilbert space decomposition $\mathscr{H}_{\rm feas} + \mathscr{H}_{\rm inf}$, which can be verified using \erfa{bin:map:op}{op:map}. This ensures there is no connection between states from the two orthogonal subspaces\footnote{Moreover, even if leakage exists (say due to implementation or Trotterization), the cost unitary only add phases to the leaked components, but it does not propagate or amplify them.}

	\section{Application to bosonic systems} \label{sec:bosonic}
	There are several models and phenomena whose behavior are purely described by boson statistics. Examples include Bose-Hubbard model (\srf{sec:bh}), magnetism \cite{SurNev24}, and boson sampling \cite{HanPan20}. 
	The Hilbert space of bosonic systems is infinite dimensional, implying that operators and basis states defined on that are not bounded. Therefore, for studying such systems on qubit-based quantum devices, the vector space needs to be truncated at a finite $D$ dimension\footnote{This also holds for other gate model platforms and/or analog quantum computing hardware to avoid the notion of infinity. From practical perspective, realizing infinite energy states is not possible.}. This poses a challenge which will be briefly explained below.
	
	Quantum harmonic oscillators conveniently model the storage and manipulation of quantum information in the Hilbert space of bosonic systems. A single bosonic mode truncated at a finite number $\mathbb{N}_c$ can be represented in the Fock basis as follows: 
	\begin{subequations}
		\begin{eqnarray}
			\ket{\Psi_0} &=& \ket{\mathbb{0}}, \quad \textrm{vacuum state} \\
			\ket{\Psi_1} &=& \ket{\mathbb{1}}, \quad \textrm{single-boson state} \\
			\vdots && \vdots \\
			\ket{\Psi_{D-1}} &=& \ket{\mathbb{N}_c}. \quad \mathbb{N}_c\textrm{-boson state}
		\end{eqnarray}
	\end{subequations}
	Here $\mathbb{N}_c = D-1$ denotes the cutoff number for the highest Fock state, indicating that the truncated Hilbert space of the oscillator is $D$-dimensional. When it comes to pick an $\mathbb{N}_c$, we should make sure the probability amplitudes of high number states remain infinitesimal throughout the calculations. 
	
	The creation and annihilation field operators acting on the Fock states are given according to
	\begin{eqnarray}
		\hat{a} \ket{\Psi_d} &=& \sqrt{d}\, \ket{\Psi_{d-1}}, \qquad d = 1,\cdots,D-1 \\
		\hat{a}\dg \ket{\Psi_d} &=& \sqrt{d+1}\, \ket{\Psi_{d+1}}, \; d = 0,\cdots,D-2,
	\end{eqnarray} 
	and $\hat{a} \ket{\Psi_0}=\hat{a}\dg \ket{\Psi_{D-1}}=0$. It is crucial to remember that in the truncated Hilbert space of the oscillator, the usual commutation relation between the bosonic operators does not hold, that is $[\hat{a},\hat{a}\dg]\neq \hat I.$ Therefore, to produce the accurate bosonic statistics in calculations, one should explicitly impose the correct algebra\ in the underlying equations containing the ladder operators. 
	
	Using different mapping strategies $m\in\{b,s,u\}$, these ladder operators transform similar to \erf{op:map}:
	\begin{equation} \label{field:multiqubit}
		\hat{a}_m = \hat{\cal M}_m \hat{a} \hat{\cal M}_m\dg = \sum_{d=1}^{D-1} \sqrt{d}\, \ket{\psi_{d-1}^K}_m\bra{\psi_d^K},
	\end{equation}
	and $\hat{a}_m\dg = \left(\hat{a}_m\right)\dg$.
	Notice that the square root factor appears in the multi-qubit representation of the raising and lowering operators. Therefore, qubit-based implementation of bosonic modes has to accommodate for its realization. Quantum circuits have been designed for this purpose, although it turns out they might be computationally expensive \cite{HanSvo18}. 
	
	After constructing the field operators, the number operator in the corresponding basis is expressed as
	\begin{equation} \label{nOp:multiqubit}
		\hat{n}_m = \hat{a}_m\dg \hat{a}_m = \sum_{d=0}^{D-1} {d}\, \ket{\psi_{d}^K}_m\bra{\psi_{d}^K}.
	\end{equation}
	Since each qudit to multi-qubit encoding make use of a different set of states, it is useful to see how number eigenstates are interpreted accordingly. 
	The binary basis can naturally represent the Fock states as each bitstring expresses an integer corresponding to the mean occupation number. Whereas the symmetric-encoded number states are distinguished by the number of excitations (or equivalently the Hamming distance with respect to the vacuum state). And for the unary encoding, the location of $1$ labels the number of quanta in Fock states.
	
	For the remaining of this section we consider applying QAOA to two bosonic systems. First, it is shown that thermal state generation can approximately be achieved via the variational procedure of QAOA. As an example, two coupled harmonic oscillators are investigated. Next, we inspect finding the approximate ground state of systems described by the repulsive Bose-Hubbard model.

	\subsection{Quantum approximate thermalization}
	Quantum thermalization is present in phenomena from particles physics to many-body systems \cite{ZhaPan22,EisGog15}. It studies the underlying processes governing the emergence of statistical mechanics from, in particular, isolated quantum systems\footnote{For open quantum systems, the state of thermal equilibrium is determined by the environment's temperature. Whereas for closed quantum systems, the eigenstate thermalization hypothesis claims entanglement spread is responsible for the information loss of an initial pure state. However, counter-examples such as integrable systems deviate from this hypothesis by showing that system evolution depends on the initial state \cite{ZhaLoh25}.} \cite{Ueda20}. Achieving perfect thermalization for practical scenarios might not be possible. For example, the integrability property implies the existence of an extensive number of conserved quantities that make certain states inaccessible. As a result, the \emph{exact} thermal distribution $\hat\rho_{\rm th}$ cannot be reached \cite{EisGog15}. Quantum approximate thermalization is an alternative way for approaching sufficiently close to thermal states. Through this process, systems initially out of equilibrium  relax towards a state resembling the Gibbs state. 
	
	Apart from interesting fundamental aspects of quantum thermalization, the latter has found applications in quantum machine learning tasks. A prominent example is the concept of quantum Boltzmann machine serving as a neural network learning model \cite{AmiMel18}. The idea here is to assume our data can be represented in a mixed-state format and use the generated thermal state $\hat\varrho_{\rm th}$ to approximate the density matrix $\hat\rho_{\rm data}$ encoding the data. To distinguish these states, a measure such as the quantum relative entropy \cite{NieChu10}
	\begin{equation} \label{eq:rel:ent}
		{\cal S}(\hat\rho_1\Vert\hat\rho_2) = {\rm Tr}[\hat\rho_1\,{\rm log}\,\hat\rho_1] - {\rm Tr} [\hat\rho_1\, {\rm log}\,\hat\rho_2], 
	\end{equation}
	may be considered. Note that the above expression can also be used to quantify the statistical distance between the true and approximate thermal states $\hat\rho_{\rm th}$ and $\hat\varrho_{\rm th}$, respectively. 
	
	It has been shown that variational quantum algorithms can be utilized for generative neural network models, creating the Gibbs state of a Hamiltonian, and in general, constructing mixed-state ansatz \cite{VerBia19,VerHid19,MarSwi19,YosFuj20}. Here, we make use of QAOA to variationally 
	approach the thermal state $\hat\rho_{\rm th} = e^{-\beta \hat{H}_C}/R_\beta$ of a problem Hamiltonian $\hat{H}_C$ where $R_\beta={\rm Tr}[e^{-\beta \hat{H}_C}]$. An approximation to the statistics of the true thermal state can be obtained by optimizing the ansatz state $\hat{\rho}(\Theta)$ over the set of parameters $\Theta$. This procedure is fulfilled through minimizing the quantum relative entropy, \erf{eq:rel:ent}, where for the target state $\hat\rho_{\rm th}$ simplifies to 
	\begin{equation} \label{eq:relent:gibbs}
		{\cal S}\left(\hat{\rho}(\Theta) \Vert \hat\rho_{\rm th}\right) = \beta\,\langle\hat{H}_C\rangle -S\left(\hat{\rho}(\Theta)\right)+{\rm log}\,R_\beta,
	\end{equation}
	with $\langle\hat{H}_C\rangle = {\rm Tr}[\hat{\rho}(\Theta) \hat{H}_C]$, and $S(\hat{\rho})=-{\rm Tr}[\hat\rho\,{\rm log}\,\hat\rho]$ is the von Neumann entropy. Note that for a fixed inverse temperature $\beta$, the minimization of relative entropy reduces to minimizing the expectation value of the cost Hamiltonian, since the von Neumann entropy remains invariant $S(\hat{\rho}) = S(\hat{U} \hat{\rho}\, \hat U\dg)$ under a unitary operation $\hat{U}$, and the last term in \erf{eq:relent:gibbs} does not depend on the ansatz. 
	
	The procedure for variational thermalization starts with preparing a rather simple Gibbs state followed by application of $p$ rounds of the QAOA layer. The idea is that the system stays close to the thermal state of the instantaneous Hamiltonian during the course of evolution until it reaches to a state approximating the thermal distribution of $\hat{H}_C$. In the asymptotic limit of large $p\rightarrow \infty$ this process is effectively simulating an adiabatic path (Trotterizing the adiabatic transformation) towards state thermalization. Quantum annealing devices can naturally simulate analog adiabatic evolution, and like other analog computing devices \cite {AghZha25,BluLuk24} can be used as Gibbs samplers \cite{VufLok22}.
	
	It should be emphasized that the generated variational state only resembles a psuedothermal state. In other words, the statistical properties of this thermal-like state is similar to that of the exact thermal state (e.g. determining expectation value of an observable). Moreover, in the ideal scenario for which QAOA consists of an infinite number of repetitions, the exact finite-temperature thermal state cannot be created \cite{VerBia19}. This is due to the design of this algorithm for finding the minimum-energy state of problem Hamiltonians. 
	
	\subsubsection{Initial state preparation} \label{sec:StdMix:RhoInit}
	In the standard QAOA with the mixing Hamiltonian $\hat{H}_M = \sum_k \hat{X}_k$, its eigenvector $\ket{-}^{\otimes K}$ is chosen as the initial state. In \srf{sec:Hmix} it was shown that, in general, the form of this Hamiltonian can be quite different than the standard choice. As a result, preparing the corresponding eigenstate might need entangling operations. However, we saw that for symmetric encoding of finite-dimensional systems the above Hamiltonian is the ideal selection. The same statement holds for binary mapping of certain problem size $D=2^K$. Therefore, here we focus our attention on these scenarios, and refer the interested reader to \arf{appn:rho:init:bin} for producing thermal states of other mixing Hamiltonians. One way to generate the initial thermal state 
	\begin{equation} \label{rho:ini:std}
		\hat{\rho}_{\rm in} = \frac{1}{R_\beta} e^{-\beta \sum_{k} \hat{X}_k}
	\end{equation}
	is inspired by quantum information theory. It is based on the fact that tracing out part of a pure entangled state leads to mixedness in the reduced state of the subsystem \footnote{For the special case of maximally entangled states, the trace operation leaves the subsystem in a maximally mixed state. The latter can be interpreted as the associated thermal state at infinite temperature ($\beta \rightarrow 0$).}. This suggests an extra register of $K$ ancillary qubits is required in addition to the $K$ qubits reserved for the problem instance. Assume the joint system is prepared in the following pure entangled state (also known as the thermofield double state in the literature \cite{WuHsi19,ZhuMon20,MatPap21})
	\begin{equation} \label{eq:rho:pa}
		\ket{\Phi}_{PA} = \frac{1}{\sqrt{R_\beta}} \sum_{\boldsymbol\jmath} e^{- \lambda_{\boldsymbol\jmath} \beta/2} \, \ket{\boldsymbol\jmath}_P \ket{\boldsymbol\jmath}_A,
	\end{equation}
	where $\ket{\boldsymbol\jmath} \equiv \ket{j_{K-1} \cdots j_1 j_0}$ with $j_k \in\{+,-\}$ is an $X$-Pauli eigenstate of $K$ qubits, the normalization factor is $R_\beta = \sum_{\boldsymbol{\jmath}} e^{- \lambda_{\boldsymbol\jmath} \beta}$, and $\lambda_{\boldsymbol\jmath} = \sum_k j_k$. Now, tracing over the ancillary degrees of freedom we obtain
	\begin{equation} \label{rho:ini:simple}
		{\rm Tr}_A\left[\,\ket{\Phi}_{PA}\bra{\Phi}\,\right] = \frac{1}{R_\beta} \sum_{\boldsymbol{\jmath}} e^{- \lambda_{\boldsymbol\jmath} \beta} \ket{\boldsymbol{\jmath}}_P\bra{\boldsymbol{\jmath}} = \rho_{\rm in},
	\end{equation} 
	which is exactly the thermal state $\rho_{\rm in}$ of the mixing Hamiltonian. 
	This can be verified by writing the spectral decomposition of $\hat{H}_M = \sum_k \hat{X}_k = \sum_{\boldsymbol{\jmath}} \lambda_{\boldsymbol\jmath} \op{\boldsymbol{\jmath}}{\boldsymbol{\jmath}}$ and substituting it in \erf{rho:ini:std} to get to the relation in \erf{rho:ini:simple}. 
	
	This procedure indicates that in order to produce the initial Gibbs state, the creation of the entangled state, \erf{eq:rho:pa}, is necessary. Thus, estimating the difficulty for generating such states could prove valuable. The required combined state is a linear superposition of GHZ-type states. Based on the discussion presented in \srf{sec:Hmix:multiq}, it is estimated that nearly $K$ entangling CNOT operations are needed to construct the joint state.

	\subsubsection{Thermalizing two coupled quantum harmonic oscillators}
	Systems of interacting resonators have proved to be useful for variety of tasks. Examples include studying the physics of quantum information \cite{YanNor16, WeaBou17}, quantum control \cite{ZhaSes25}, and quantum thermodynamics \cite{WanHe15,SheWu21}. Here, we investigate approximate thermalization of two intermediately coupled quantum harmonic oscillators described by the following Hamiltonian ($\hbar = 1$):
	\begin{equation} \label{Ham:QHO:gen}
		\hat H_{\rm CHO} = \omega_1 \hat n_1 + \omega_2 \hat n_2 + \lambda (\hat{a}_2 \hat{a}_1\dg + \hat{a}_2\dg \hat{a}_1),
	\end{equation}
	where $\hat{n}_\ell = \hat{a}_\ell\dg \hat{a}_\ell$ is the number operator for the $\ell-$th oscillator, $\omega_\ell$ the corresponding angular frequency, and $\lambda$ denotes the coupling strength. This example allows us having access to the exact thermal state of the system for benchmarking the QAOA performance. Obviously, when there is no interaction between the oscillators, the thermal state is simply the product of individual thermal distributions $\hat{\rho}_{\rm th}^{(j)} = \sum_{n_j} p_{n_j} \op{n_j}{n_j}$ where $p_{n_j} = e^{-\beta n_j \omega_j}/\sum_{n_j} e^{-\beta n_j}$. In the presence of coupling, the spectral decomposition of $\hat{H}_{\rm CHO} = \sum_k \varepsilon_k \op{\varepsilon_k}{\varepsilon_k}$ should be sought to determine its Gibbs state (consult \arf{appn:spec:Hqho} for the details)
	\begin{equation} \label{rho:th:exact:gen}
		\hat{\rho}_{\rm th} = \frac{\sum_k e^{-\beta \varepsilon_k}\op{\varepsilon_k}{\varepsilon_k}}{\sum_k e^{-\beta \varepsilon_k}}.
	\end{equation}
	
	\begin{figure} 
		\includegraphics[width=\linewidth]{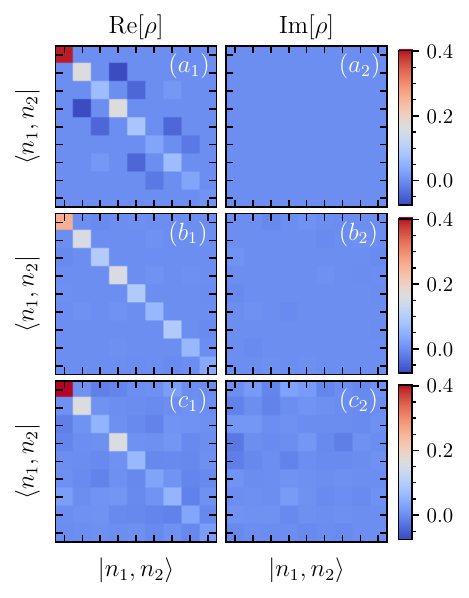}
		\caption{\label{fig:qat:rho} (Color online). Approximate thermalization of two coupled harmonic oscillators. (a) Top row is the exact thermal density matrix (real and imaginary parts) obtained via \erf{rho:th:exact:gen}. (b) The middle and (c) bottom rows show the simulation results using the binary $\rho{(\Theta_b^*)}$, and symmetric $\rho{(\Theta_s^*)}$ encodings, respectively. The quantum relative entropy, \erf{eq:rel:ent}, for these mappings are ${\cal S}(\rho{(\Theta_b^*)}|\rho_{\rm th}) = 0.26$  and ${\cal S}(\rho{(\Theta_s^*)}|\rho_{\rm th}) = 0.14$, and the respective state fidelities, \erf{mixed:fidelity}, are ${\cal F}(\rho{(\Theta_b^*)},\rho_{\rm th}) = 0.89$, and  ${\cal F}(\rho{(\Theta_s^*)},\rho_{\rm th}) = 0.93$. Here $\Theta_m^*$ are optimal circuit parameters for mapping scheme '{\emph m}' with $p=5$ QAOA layers. Consult the main text for the effect of noisy (depolarized) CNOT gate on the algorithm's efficiency.  We set $\omega_1 = \omega_2 = 2$, $\lambda = 1$, $\beta = 0.5$, and with $10^3$ shots. Classical parameter optimization is performed using the COBYQA algorithm.
		}
	\end{figure}
	
	The total mean bosonic excitations $N=\sum_\ell \an{\hat{n}_\ell}$ is not a constant of motion since it does not commute with both $\hat{H}_{\rm CHO}$ and $\hat{H}_{M,m}$. Therefore, the choice of problem's parameter regime should be such that on average the total number of bosons does not exceed the cutoff number ${\mathbb N}_c$ at which the Fock space is truncated. We assume there are in total two bosons in the system such that the highest number state has $\mathbb{N}_c = 2$ excitations. Therefore, each resonator is effectively approximated by a qutrit ($D=3$). This means either of the binary or symmetric encoding requires only $K=2$ qubits to map such subsystem. The latter is represented by \erfa{ket:map:b:d3}{ket:map:s:d3}, for the respective encoding schemes with $\ket{\Psi_d} \in \{\ket{\mathbb{0}}, \ket{\mathbb{1}}, \ket{\mathbb{2}}\}$. As a result, the joint state $\ket{\Psi_d,\Psi_{d'}}$ of the system in the original space is mapped onto $\ket{\psi_d,\psi_{d'}}$ in the multi-qubit space. For example, the single-boson state $\ket{\mathbb{0},\mathbb{1}}$ is represented by $\ket{0001}$ and $\frac{1}{\sqrt{2}}(\ket{0001}+\ket{0010})$ using the binary and symmetric encoding, respectively. 
	
	Let us now express the problem Hamiltonian in the multi-qubit Hilbert space. In the binary scheme the field operator, \erf{field:multiqubit}, takes the following form 
	\begin{eqnarray} \label{aOp:bin:d3}
		\hat{a}_b &=& \frac{1}{4}\big[IX+ZX+\sqrt{2}\,(XX+YY) \nonumber \\
		& &+\,i\big(IY+ZY+\sqrt{2}\,[YX-XY]\big)\big],
	\end{eqnarray}
	and the number operator given in \erf{nOp:multiqubit} reads
	\begin{equation} \label{nOp:bin:d3}
		\hat{n}_b = \frac{1}{4} \left(3II + IZ -ZI - 3ZZ\right).
	\end{equation}
	Once these operators are determined, it is easy to construct the Hamiltonian. For example, the first term transforms according to $\hat{n}_1 \to II\otimes \hat{n}_b$, whereas the second one converts to $\hat{n}_2 \to \hat{n}_b \otimes II$. The same procedure applies to the interaction expression in \erf{Ham:QHO:gen}.

	For $D=3$ dimensional systems the qubit count in the symmetric encoding is the same as in the binary mapping. Nevertheless, the distinct aspect of the symmetric basis (in that it requires entangling manipulation of some eigenbases) brings about more complexity when it comes to translate the problem Hamiltonian to a quantum circuit. 
	Now, using \erf{ket:map:s:d3}, \erfa{nOp:multiqubit}{field:multiqubit} the field and number operators are presented as the following:
	\begin{eqnarray}
		\hat{a}_s &=& \alpha_+ (IX + XI + ZX + XZ) \nonumber \\
		& & +\,i\, \alpha_- (IY + YI + ZY + YZ)
	\end{eqnarray}
	where $\alpha_\pm = ({1\pm \sqrt{2}})/{4\sqrt{2}}$, and
	\begin{equation} \label{nOp:sym:d3}
		\hat{n}_s = \frac{1}{4} \left(3II -2IZ - 2ZI + ZZ + XX + YY\right).
	\end{equation}
	For constructing the objective Hamiltonian the same line of argument described after \erf{nOp:bin:d3} holds here. Equipped with these operators, the unitary evolution generated by the problem Hamiltonian can be established. Knowing the initial state of the variational circuit and the mixer, we are set to execute different layers of the QAOA.

	Figure \ref{fig:qat:rho} illustrates the exact and approximate thermal states obtained via simulations for two interacting resonators, \erf{Ham:QHO:gen}. The top row depicts the real and imaginary parts of the ideal Gibbs state determined through diagonalization of the coupled harmonic oscillators (see \arf{appn:spec:Hqho} for the details of calculations). The outcome of numerical simulations using binary encoding is shown in the middle panel (in this work we mainly use the constrained optimization by quadratic approximation (COBYQA) method as the optimizer). Here we employ $H_{M,b}^{(2)}$, \erf{H:mix:d3s2}, as the mixing Hamiltonian as it yields less quantum relative entropy ${\cal S}(\rho{(\Theta_b^*)}|\rho_{\rm th}) = 0.26$ compared to other choices \erf{H:mix:d3s1} and \erf{H:mix:d3s3} or even the combined selection \erf{mixer:gen}. Note that $\Theta_m^*$ is the optimal set of parameters for the mapping scheme \emph{m}. We also check the fidelity between a reference state $\hat{\varrho}$ and the reconstructed one $\hat{\rho}$ defined according to \cite{Joz94}
	\begin{equation} \label{mixed:fidelity}
		{\cal F}(\hat\rho,\hat\varrho) = \left({\rm Tr}\left[\sqrt{\sqrt{\hat\rho}\, \hat\varrho \,\sqrt{\hat\rho}}\right]\right)^2.
	\end{equation}
	For the scenario considered above, we obtain ${\cal F}(\rho{(\Theta_b^*)},\rho_{\rm th}) = 0.89$. The symmetric-encoded simulations show better performance by further boosting up state fidelity ${\cal F}(\rho{(\Theta_s^*)},\rho_{\rm th}) = 0.93$ with respect to the exact thermal state, and decreasing quantum relative entropy ${\cal S}(\rho{(\Theta_s^*)}|\rho_{\rm th}) = 0.14$. Here we fix angular frequencies $\omega_1 = \omega_2 = 2$, the coupling strength $\lambda=1$, inverse temperature $\beta=0.5$, and $p=5$. Therefore, the binary mapped mixing Hamiltonian requires $2p=10$ CNOT gates versus none for the symmetric encoding. Even taking into account the entangling measurement, which we estimate it would need two CNOT gates (one for each Bell measurement), the symmetric encoding is still less costly from implementation viewpoint \footnote{Assuming that the state preparation for both strategies would require more or less the same number of CNOT gates (see \srf{sec:StdMix:RhoInit} and \arf{appn:rho:init:bin})}. 
	
	\begin{figure} 
		\includegraphics[width=\linewidth]{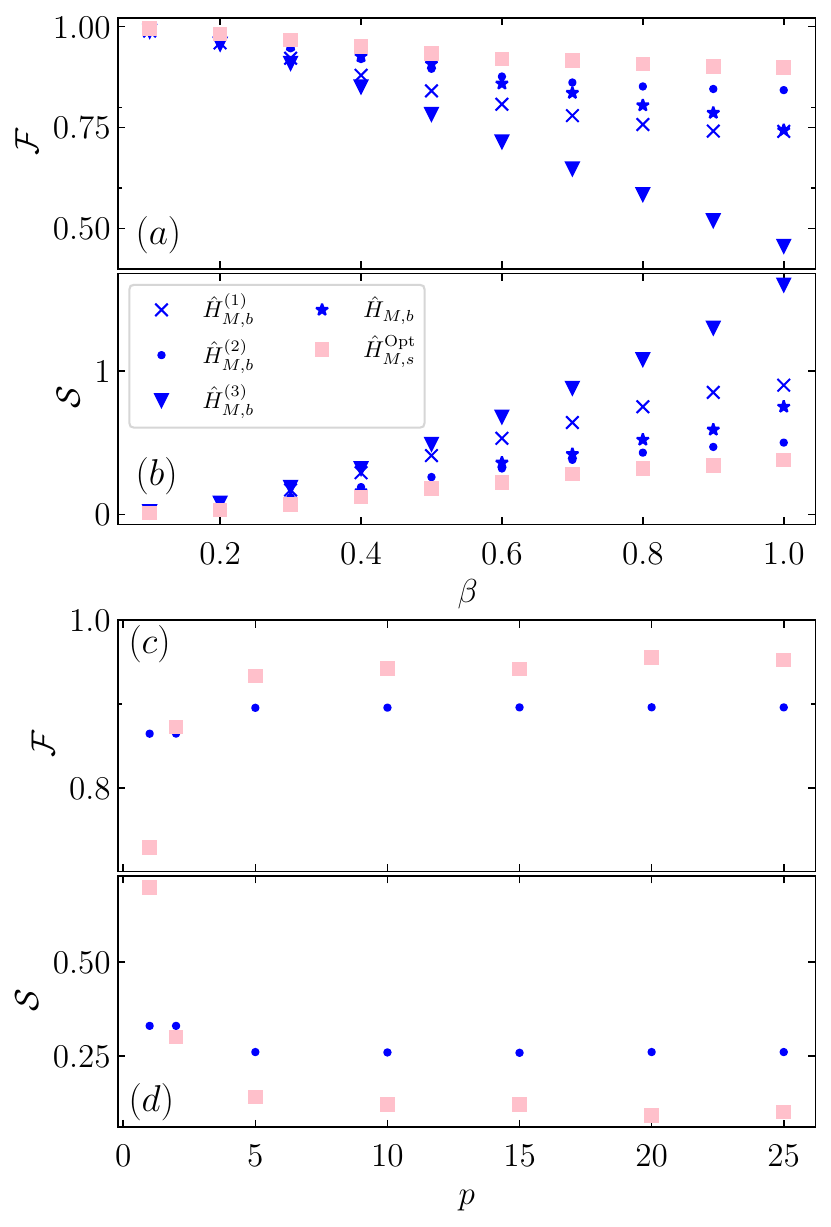}
		\caption{\label{fig:fident} (Color online). Quantifying approximate thermalization. (a) and (b) demonstrate variations of fidelity and quantum relative entropy in terms of inverse temperature. Simulations show that symmetric encoding (pink squares) is more robust to temperature changes compared to the standard mapping. For binary encoding the choice of mixing Hamiltonian is going to impact accuracy of computations. For example, the $XY$ driver Hamiltonian has the worst performance among all other options. (c) and (d) depict the same metrics as a function of the number $p$ of QAOA layers. The profiles suggest that as the circuit depth increases the classical optimization methods struggle with finding the optimal solution. Here we set $\omega_1 = \omega_2 = 2$, $\lambda = 1$, and the number of shots to $10^3$; for (a) and (b) $p=5$ and for (c) and (d) $\beta = 0.5$. We employ the COBYQA as the classical optimizer.
		}
	\end{figure}
	
	We also impose depolarizing noise with probability in the interval $p_d \in [0.02,0.05]$ to the controlled-NOT operation to see its impact on the protocol's performance\footnote{The two-qubit depolarizing channel is modeled as \cite{NieChu10}: $$ {\cal E}(\rho) = (1-p_d) \rho + \frac{p_d}{15} \sum_{\sigma \neq I\otimes I} \hat \sigma \rho\, \hat \sigma, $$ where $\hat \sigma \in \{I, X, Y, Z\}^{\otimes2}$, and $0 \le p_d \le 1$ denotes the depolarizing probability. The effect of this channel on a CNOT gate is modeled by first applying the ideal entangling operation $\rho \mapsto \tilde{\rho} \equiv \hat U_{\rm CNOT}\, \rho\, \hat U_{\rm CNOT}\dg$ followed by the depolarizing map ${\cal E}(\tilde{\rho})$.}, although it has been shown that this is the most tolerant two-qubit gate against such noise \cite{HarNie03}. As expected, the presence of depolarizing error channel reduces the state fidelity to ${\cal F}_{\rm dn}(\rho{(\Theta_b^*)},\rho_{\rm th}) \approx [0.86,0.85]$ and ${\cal F}_{\rm dn}(\rho{(\Theta_s^*)},\rho_{\rm th}) \approx [0.92,0.91]$. Consequently, the quantum relative entropy grows by a small amount within ${\cal S}_{\rm dn}(\rho{(\Theta_b^*)}|\rho_{\rm th}) \approx [0.35,0.36]$, and ${\cal S}_{\rm dn}(\rho{(\Theta_s^*)}|\rho_{\rm th}) \approx [0.18,0.21]$, respectively. It would be interesting to consider the effect of other sources of noise on the protocol, which we leave it for future work.
	
	To reiterate, these psuedothermal states can serve for approximate sampling of the exact thermal distribution in order to calculate expectation value of an observable. For example, let us consider the mean occupation number $\an{\hat{n}_j}$ for each oscillator. The analytical solution, \erf{rho:th:exact:gen}, provides a reference value $0.48$, while the binary and symmetric mapping techniques result in $\approx 0.68$, and $\approx 0.45$, respectively. Although the quality of generated states partially justifies these figures, the underlying reason can be related to the initial state. The latter is prepared at the same inverse temperature as that of the target state, however the Fock state populations differ depending on the mixing Hamiltonian. Consequently, the starting average number of bosons is different. In addition, initially the infeasible states may have nonzero amplitude and due to restricting the dynamics to the feasible subspace they cannot contribute to mean excitation number. In the following we will present another reason in relation to difficulty of producing low-temperature Gibbs states that sheds light on the above observation.

	Figure \ref{fig:fident} demonstrates the performance of the quantum approximate thermalization algorithm by varying the inverse temperature and number of QAOA layers. In \frf{fig:fident}(a) and (b) fidelity, \erf{mixed:fidelity}, and quantum relative entropy, \erf{eq:rel:ent}, with respect to the target thermal state as a function of $\beta$ are shown. Irrespective of the encoding technique, it can be seen that as $\beta$ increases the approximated thermal state has lower overlap with the Gibbs state of the problem Hamiltonian. The reason for degradation in fidelity, and hence increase in relative entropy, is rooted in the concentration of states near low-energy region of the thermal distribution at lower temperatures. Since the initial thermal state distributes almost evenly across the energy spectrum,
	this intuitively implies a rather less efficient heat transfer process due to lack of decent temperature gradient. In contrast, at higher temperature (small $\beta$'s) the desired Gibbs state tend to populate the Fock eigenstates with comparable probabilities. In other words, the corresponding density matrix is equivalent to that of a mixed state. Therefore, the initial thermal state facilitates enough heat flow within the system until it approaches the thermal equilibrium \footnote{Analysis of the total mean boson number $N$ also provides an insight to interpret this behavior. For low temperature thermalization the value of $N$ has a bigger gap between the target and initial states, whereas this gap is less significant for smaller $\beta$'s. In both scenarios the dynamic starts at a higher number of bosons than the final state.}.
	
	Comparing the two mapping schemes, overall the symmetric encoding shows better performance than the binary one. Among different binary driving Hamiltonians, $\hat{H}_{M,b}^{(2)}$ given in \erf{H:mix:d3s2} has the best execution and the pure $XY$ mixer, \erf{H:mix:d3s3}, exhibits poor behavior. 
	
	The outcome of examining different ansatz on the variational algorithm is presented in \frf{fig:fident}(b). It might be thought that by hiking the number $p$ of the QAOA layers the quality of candidate solutions will improve. Even though, there is certainly enhancement in the metrics such as the state fidelity, or equivalently decline in the quantum relative entropy, beyond some values the gain is not significant. In particular, if we take into account the implementation cost, it would be hard to justify ansatz with a larger circuit depth. Incremental achievement could be well related to the Barren plateau and the difficulty of classical optimizer routines in searching the optimization landscape for finding the optimal or near-optimal set of parameters \cite{ArrCol21,LarCer25}. Note that the symmetric encoding outperforms the binary mapping for $p>1$. The fact that it also requires less CNOT gates is beneficial so long realization is concerned.     
	
	\subsection{Many finite-level systems: the Bose-Hubbard model} \label{sec:bh}
	In this section our goal is to examine the application of QAOA for finding the ground state of systems that are made up of $L$ subsystems each of dimension $D$. Due to the exponential growth of the multiqubit's Hilbert space dimension, it is obvious that the infeasible subspace is much larger than that of the allowed dynamical events. In this instance, adding a penalty term to the objective function to avoid illegitimate states might not be an efficient strategy. 
	
	\begin{figure*}
		\centering
		\subfloat
		{\includegraphics[width=0.45\textwidth]{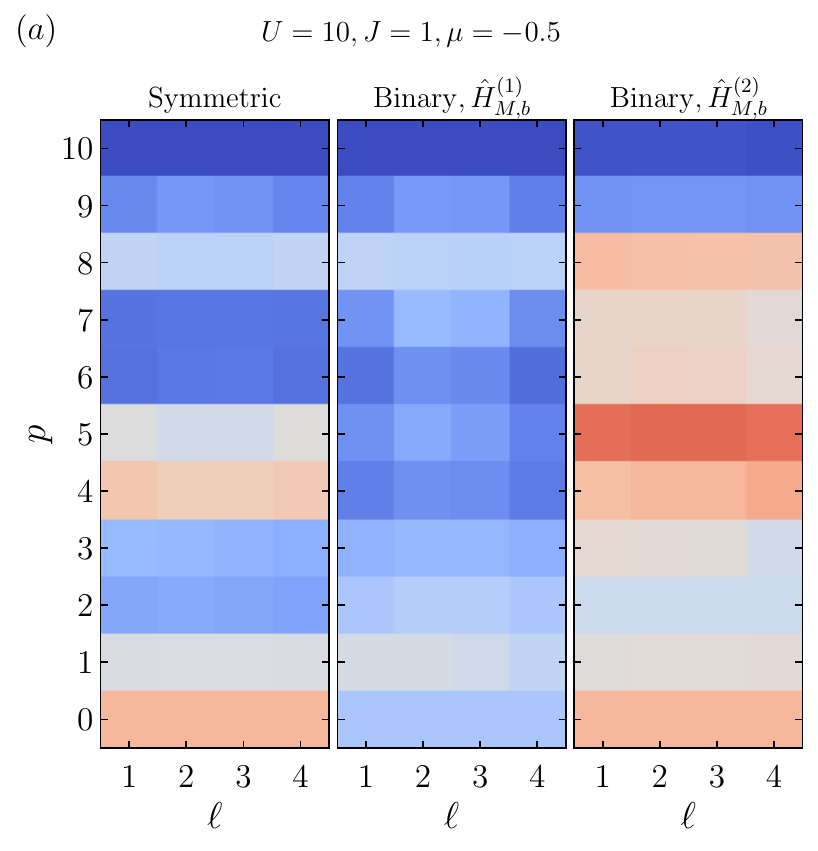} \label{fig:bh:u10}}
		\hfill 
		\subfloat{\includegraphics[width=0.525\textwidth]{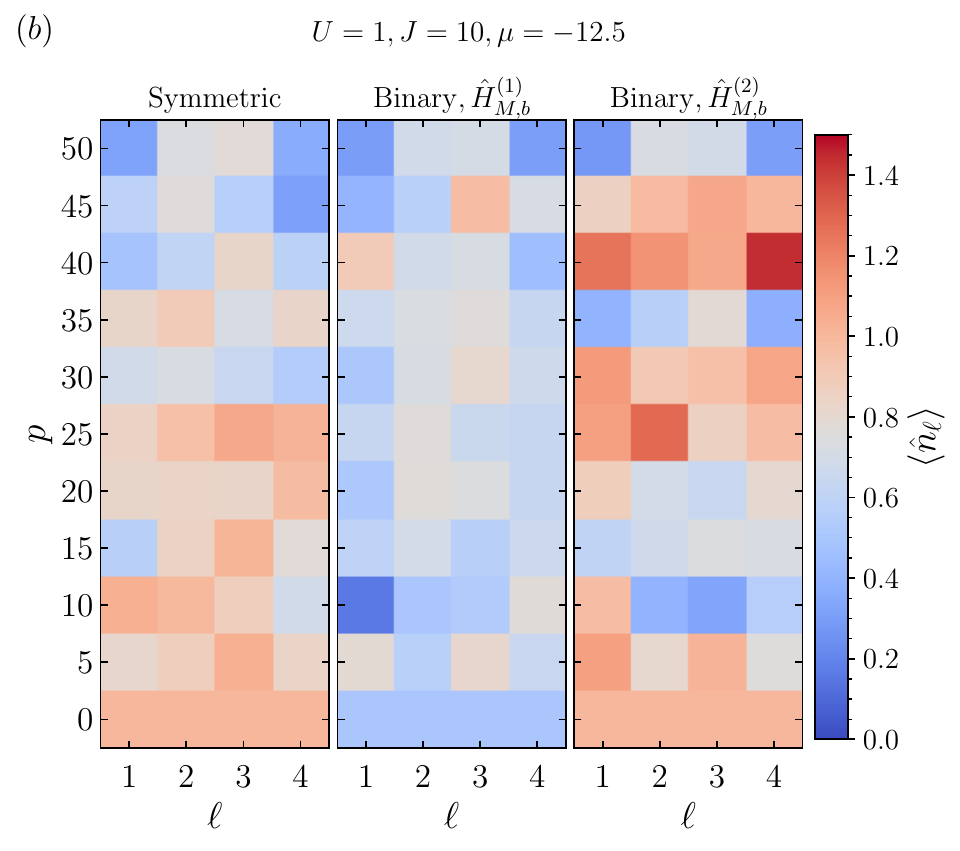} \label{fig:bh:u1}}
		\caption{(Color online). Distribution of boson number per site in the (a) strong and (b) weak interaction regime for symmetric and binary encoding schemes. For the latter the system is initialized in the ground state of the respective mixing Hamiltonian, built upon \erfa{H:mix:d3s1}{H:mix:d3s2}. The corresponding initial mean particle number is indicated by $p=0$. For the interaction dominated case, both mappings converge to the average occupation number $\langle \hat{n}_\ell \rangle$ obtained over the ground state $\ket{\Psi_{\rm gr}}$ of $\hat{H}_{\rm BH}$ (for the chosen parameters, this is the vacuum state $\ket{{\mathbb 0}{\mathbb 0}{\mathbb 0}{\mathbb 0}}$). The fidelity with respect to this minimum-energy eigenstate is ${\cal F}_m \ge 0.95$. In contrast, the ground state of the kinetically dominated scenario is quantum correlated. Therefore, an ansatz circuit with a higher depth, for example $p=50$, is required to represent a larger range of quantum states. This implies the classical optimizer needs more resources. With these parameters, the final state reaches ${\cal F}_{b} \approx (0.92,0.99)$ for the corresponding mixer $\hat{H}_{M,b}^{(j)}$, and ${\cal F}_s \approx 0.84$. The target mean particle number is $\langle \hat{n}_\ell \rangle \approx (0.28,0.72,0.72,0.28)$.}
		\label{fig:bh}
	\end{figure*}
	
	Here as an example we consider the Bose-Hubbard model which describes correlated bosonic particles in a lattice potential. This formalism is well-known for capturing the physics of superfluid to Mott-insulator phase transition. It is usually realized in ultracold atoms trapped in optical lattices \cite{GreBlo02}. These setups can serve as quantum simulators to study complex many-body systems \cite{ChrBlo17, SchTak20}. However, analog quantum simulators face challenges such as not being able to realize certain Hamiltonian terms. 
	Now, consider the following Bose-Hubbard Hamiltonian \cite{FisFis89}    
	\begin{equation} \label{Ham:BH:gen}
		\hat{H}_{\rm BH} = \hat{H}_J + \hat{H}_0,
	\end{equation}
	where 
	\begin{subequations}
		\begin{align}
			\hat{H}_0 &= - \mu \sum_{\ell=1}^{L} \hat{n}_\ell + \frac{U}{2} \sum_{\ell=1}^{L} \hat{n}_\ell (\hat{n}_\ell - 1),  \label{Ham:BH:local} \\
			\hat{H}_J &= -J \sum_{\ell=1}^{L-1} \left(\hat{a}^\dagger_\ell \hat{a}_{\ell+1}^{} + {\rm h.c.}\right). \label{Ham:BH:hop}
		\end{align}
	\end{subequations}
	
	Here $L$ is the number of sites, $\mu$ the chemical potential, $U$ denotes on-site interactions, and $J$ represents the hopping strength. These competing parameters determine whether the system is in either of superfluid or insulator quantum phases or in a transition process between the two. On the one hand, repulsive interactions ($U>0$) tend to steer the system in the vacuum state, whereas attractive interactions ($U<0$) give rise to clustering of bosons. On the other hand, the hopping term is responsible for tunneling of bosons through nearest-neighbor potential barriers. The energy cost of changing the total boson number by one is associated with the chemical potential.
	
	Time evolution of the system via the propagator ${\rm Exp}({-i\hat{H}_{\rm BH} t})$ preserves the total number of bosons. However, in QAOA trial states are generated through alternative applications of $\hat{U}_M(\nu)$ and $\hat{U}_C(\gamma)$. Since $\hat N$ and $\hat{H}_M$ do not commute, then the total occupation number is not conserved under the unitary evolution generated by the mixer. Notably, the form of the mixing Hamiltonian  dictates the initial mean boson number in each site\footnote{Note that, the problem may effectively be considered as an open quantum system in which through interaction with an environment bosonic particles are added or removed with two caveats. Firstly, the total number of bosons should change within the limitations of truncated Hilbert space. Secondly, the simulations are not dynamical in the sense that an initial state is refined until it converges to the target state through variational processes.}. 
	
	With this in mind our goal is to find the minimum-energy eigenstate of the BH Hamiltonian with only repulsive interactions. We start with making use of the same field and number operators presented in \erfa{aOp:bin:d3}{nOp:sym:d3}  to transform the BH Hamiltonian to its multi-qubit representation. Besides, one should be cautious to make sure that field operators are normally ordered because $\hat{M} \hat{M}\dg\neq I_{\rm qb}$. Therefore, the second summation in \erf{Ham:BH:local} can be rearranged to $\sum_\ell \hat{a}_\ell\dg \hat{a}_\ell\dg \hat{a}_\ell^{} \hat{a}_\ell^{}$.
		
	We can gain insight into choosing suitable parameters by considering simplified cases. In the absence of hopping $J = 0$, the system is solely described by the so-called local Hamiltonian $\hat{H}_0$. The characteristic of this insulating phase is that each cell has a well-defined particle number (the Hamiltonian becomes diagonal in the number basis). In this strong interaction regime the ground state is in a tensor product form $\ket{\Psi_{\rm gr}(J=0)} = \ket{n_{\rm gr}}^{\otimes L}$ with $n_{\rm gr}$ bosons per site for $(n_{\rm gr}-1) < \mu/U < n_{\rm gr}$. This indicates that for the cutoff occupation number ${\mathbb N}_c = 2$, there is an upper bound on chemical potential $\mu_c<2 U$ (beyond this critical value, the truncated state space does not fully account for all valid eigenstates). In this scenario even if sites are filled with ${\mathbb N}_c$ bosons, since particles are localized to sites, then the possibility of populating higher Fock states is diminished. 
	
	Figure \subref*{fig:bh:u10} demonstrates site occupancy for a $10$-layer QAOA circuit. The Bose-Hubbard system has $L = 4$ lattice sites, with Hamiltonian parameters $U=10$, $J=1$, and $\mu = -0.5$. This is an interaction dominated regime with the vacuum state as its ground state. The initial boson distribution, displayed by $p = 0$, is dictated by the ground state of the chosen mixing Hamiltonian. It can be seen how the boson distribution changes after each round until it approximately converges to $n_{\rm gr} = \sum_\ell \langle \hat{n}_\ell \rangle = 0$ for both symmetric (left column) and binary (middle and right columns) encoding schemes\footnote{We note that the optimizer (Constraint Optimization by Quadratic Approximation) runs over the entire $p=10$ rounds. Assume we execute the classical optimization routine individually for each $p<10$. Thus, due to different optimization landscapes (or distinct parameterized quantum circuit), the resulting boson distribution would be different than what reported in \frf{fig:bh}.}. To quantify the algorithm's performance we employ the fidelity metric. Rewriting \erf{mixed:fidelity} for pure states one obtains:
	\begin{equation} \label{fid:pure:bh}
		{\cal F}_m \coloneq |\langle\Psi(\Theta_m^*)| \Psi_{\rm gr} \rangle|^2, \qquad m \in \{b,s\}.
	\end{equation}
	After plugging in the relevant parameters, near-optimal states with fidelity ${\cal F}_m > 0.99$ are obtained for the pair of mapping methods (for the binary encoding using $\hat{H}_{M,b}^{(2)}$ this figure is around $0.95$). Should we set $p=2$, the state fidelity reads ${\cal F}_m \approx 0.94$, irrespective of mapping. We also note that the choice of $XY$ mixer, \erf{H:mix:d3s3}, exhibits poor performance.   
	
	In the kinetically dominated regime ($U = 0$), particles relax in the eigenstate of lowest kinetic energy. This ground state, up to a normalization factor,
	is proportional to a coherent superposition of all eigenkets with exactly $N$ bosons $\ket{\Psi_{\rm gr}(U=0)} \propto (\frac{1}{\sqrt{L}} \sum_\ell \hat{a}_\ell\dg)^N \ket{{\mathbb 0}}^{\otimes L}$ \cite{Zwe03}. In this superfluid phase bosons can hop from one site to its neighboring one, they become delocalized and spread across the sites. Therefore, this analysis suggests within the truncated Hilbert space at most two particles can exist over the $L=4$ lattice sites.
	
	The results of our simulations in this regime with $J = 10$, $U = 1$, and $\mu = -12.5$ (this choice of chemical potential ensures $\sum_\ell \langle \hat{n}_\ell \rangle = 2$) demonstrate that both encoding approaches converge to the average occupation number $\bra{\Psi_{\rm gr}} \hat{n}_\ell \ket{\Psi_{\rm gr}} \approx (0.28,0.72,0.72,0.28)$, see Fig.~\subref*{fig:bh:u1}. However, this comes at the cost of increasing the circuit depth and demanding more classical resources\footnote{Increasing $p$ makes the optimization landscape more complex for the classical solver \cite{BitKli21}. Perhaps a global optimizer might be able to boost up the performance of COBYQA and alike local optimizers tested here \cite{GleRos24}.}. In other words, the expressibility of the ansatz becomes critical for generating the complex entangled ground state. This is in stark contrast to the strong interaction regime for which the ground state is in the product form. Setting $p=50$, the state fidelity, \erf{fid:pure:bh}, with respect to the entangled target state is $(0.92, 0.99)$ and $0.84$, for the corresponding binary and symmetric mapping techniques, respectively. The latter figure can be improved, for example, by increasing the tolerance of COBYQA optimizer\footnote{ We emphasize that these results are not intended as a benchmark of classical optimization performance, but rather as illustrative demonstrations of the circuit constructions; optimization challenges are discussed only to contextualize the numerical behavior observed.} (or adding more layers, say $p=60$, which boosts the fidelity to ${\cal F}_m \approx 0.92$). 
	
	\section{Conclusion}   \label{sec:conclusion}
	Quantum simulation and approximate optimization of many-state systems on qubit-based quantum hardware require mapping from the $
	D$-dimensional to multi-qubit Hilbert space. Depending on the encoding scheme employed, the infeasible subspace varies in size. One way to exclude illegitimate states is through penalizing the objective function. However, as the system's dimension grows, searching the vast infeasible configuration space can become inefficient. 
	
	In this work we tackle this issue within the framework of quantum approximate optimization paradigm.  It is shown that by carefully designing the mixing Hamiltonian in QAOA, the dynamics will be restricted to the feasible subspace. We examine this idea using binary, symmetric and unary encoding methods. In particular, we estimate the CNOT gate count for generating the initial state, implementing the mixer and final measurements. 
	
	It becomes evident that the standard choice of driving Hamiltonian (sum of the $X$ Pauli operators) is the best option for symmetric encoding as it only needs the Hadamard operation which can effectively be counted as free. The hardware efficient binary mapping demands less qubit resources compared with the other encoding techniques. Although this implies smaller infeasible subspace dimension, the fact that the corresponding mixing Hamiltonian necessitates realizing CNOT gates can negate that advantage (especially that there is a $p$-fold increase in that figure for the $p$-layer QAOA). 
	
	Several potential areas remain for future exploration. Firstly, as quantum hardware matures, robust QAOA should be developed to incorporate error mitigation and/or correction \cite{HePis25, OmaPis25}. The large infeasible subspace available in the symmetric and unary mapping may be exploited towards this goal. Secondly, qudit-based quantum architectures have recently emerged as a promising candidate to tackle high-dimensional problems \cite{WanKai20,DelKas23, KimLim24}. It would be worthwhile to evaluate their performance despite the fact that qudit gate fidelity might not be high enough. Thirdly, even though we considered depolarizing noise for the CNOT gate (in quantum approximate thermalization), it would be interesting to investigate the impact of other experimental imperfection by emulators and running simulations on an actual quantum device \cite{StiGar21}. Fourthly, the raw entangling gate count can be contextualized in the presence of implementation factors, alongside evaluating the required classical resources, including the effects of initialization strategies, optimizer choice, and gradient-based training approaches. Finally, it would be an interesting direction to explore the usage of a family of binary-encoded mixing Hamiltonians either by alternating or stochastic selection for each layer of QAOA. That is to say, instead of keep applying a fixed driver for the entire algorithm, one decides whether to cycle through the partial mixers or randomly choose them.

	\section*{Acknowledgment}
	The author would like to thank Laura Cunningham for reading the manuscript. We acknowledge Curtin University Digital \& Technology Solutions for providing cloud computing and storage resources. The author acknowledges the anonymous referee for constructive comments that improved the clarity and scope of this work.

	\appendix
	\numberwithin{equation}{section}
	

	\section{Initial state for binary mapping ($D=3$)} \label{appn:rho:init:bin}
	As it was mentioned in the main text, quantum circuits for approximate thermalization process require to be initialized in the thermal state of a selected mixing Hamiltonian. Since the standard mixer is not suitable for the binary encoding of systems with $D \neq 2^K$, then the exact steps presented in \srf{sec:StdMix:RhoInit} cannot be employed. Here, as an example, we consider preparation of the thermal state associated with
	\begin{equation} \label{Hmix:ham1:gen}
		\hat{H}_{M,b}^{\rm CHO} = II \otimes \hat{H}_{M,b}^{(1)} + \hat{H}_{M,b}^{(1)} \otimes II,
	\end{equation} 
	where $\hat{H}_{M,b}^{(1)} = \smallfrac{1}{2}(IX+ZX)$, \erf{H:mix:d3s1}.  The procedure is similar to that of \srf{sec:StdMix:RhoInit}; through purification we start with a joint problem plus ancillary state $\ket{\Phi}$, and by tracing out the ancillary qubits the desired state is created. To this end, let us first determine the spectral decomposition of $\hat{H}_{M,b}^{\rm CHO}$ resulting in the below eigenspace
	\begin{equation}
		\varepsilon \in \{-2,-1_4, 0_6, +1_4, 2\},
	\end{equation} 
	where subindices represent multiplicity in eigenvalues degeneracy. The respective eigenstates are
	\begin{align}
		\ket{\varepsilon} \in &\; \{\ket{-0-0}, \ket{01-0}, \ket{11-0}, \ket{-011}, \ket{-001}, \nonumber \\
		&\quad 	 \ket{0101}, \ket{\varepsilon_-}, \ket{\varepsilon_+},\ket{1101}, \ket{1111}, \ket{\varepsilon_{11}}, \\
		& \quad \ket{+001}, \ket{+011}, \ket{11+0}, \ket{01+0}, \ket{+0+0} \}, \nonumber
	\end{align}
	where $\ket{\varepsilon_\pm} = \smallfrac{1}{2}(\ket{1000}-\ket{0010})\pm \smallfrac{1}{\sqrt{2}} \ket{0111}, \ket{\varepsilon_{11}}=\smallfrac{1}{\sqrt{2}} (\ket{0000}+\ket{1010})$, and $\ket{\pm} = \smallfrac{1}{\sqrt{2}} (\ket{0}\pm \ket{1})$ are eigenkets of $X$ Pauli matrix. Therefore, the thermal state $\hat{\rho}_{\rm in}$ can be obtained via plugging in the above relations in \erf{rho:th:exact:gen}. The compound system-ancilla state should be prepared in 
	\begin{equation}
		\ket{\Phi} = \sum_{k=0}^{2^4-1} \sqrt{\Lambda_k}\, \ket{\phi_k}  \ket{\varepsilon_k},
	\end{equation}
	where $\{\ket{\phi_k}\}$ is an orthonormal basis for the auxiliary subsystem, and $\Lambda_k = e^{-\beta \varepsilon_k}/\sum_k e^{-\beta \varepsilon_k}= e^{-\beta \varepsilon_k}/[2 \,{\rm cosh}(2\beta) + 8 \,{\rm cosh(\beta)}+6]$. Now, removing the ancilla's degrees of freedom we obtain
	\begin{equation}
		\hat{\rho}_{\rm in} = {\rm Tr}_A\left[\op{\Phi}{\Phi}\right]. 
	\end{equation}
	To generate $\ket{\Phi}$ quantum circuit synthesis process may be utilized. In hindsight, we know that the symmetric encoding carries out quantum approximate thermalization task effectively compared to the binary mapping. Therefore, in this particular scenario, we do not delve into detail analysis of the circuit design. For mere benchmarking purpose, quantum simulators allow for feeding in an arbitrary initial state.
	
	For the Bose-Hubbard problem we only need to initialize in the ground state $\ket{\varepsilon_0}_{\rm BH}$  of
	\begin{align} \label{Hmix:ham1:gen:bh}
		\hat{H}_{M,b}^{\rm BH} &=  I^{\otimes 6} \otimes \hat{H}_{M,b}^{(1)} + I^{\otimes 4} \otimes \hat{H}_{M,b}^{(1)} \otimes I^{\otimes 2} \nonumber \\
		& +  I^{\otimes 2} \otimes \hat{H}_{M,b}^{(1)} \otimes I^{\otimes 4} + \hat{H}_{M,b}^{(1)} \otimes I^{\otimes 6},
	\end{align} 
	where $I^{\otimes k}$ denotes the $2^k$-dimensional identity operator. Since each site is modeled with two qubits, then the minimum-energy state of the system composed of 4 sites is a coherent linear superposition of 8-qubit basis states. If other mixing Hamiltonians are employed in \erf{Hmix:ham1:gen} and/or \erf{Hmix:ham1:gen:bh}, the above relationships for initial states should be updated accordingly.

	\section{Diagonalization of the coupled resonators Hamiltonian} \label{appn:spec:Hqho}
	Assuming equal frequencies $\omega_1=\omega_2 = \omega$ for the coupled harmonic oscillators described by the Hamiltonian given in \erf{Ham:QHO:gen}, then diagonalizing the latter leads to the following eigenvalues
	\begin{align}
		\varepsilon_0 &= 0, & \varepsilon_1 &= -2(\lambda-\omega), & \varepsilon_2 &= 2\omega,  \nonumber \\
		\varepsilon_3 &= 4\omega, & \varepsilon_4 &= 2(\lambda+\omega), &  \varepsilon_5 &= -\lambda+\omega, \nonumber \\
		\varepsilon_6 &= \lambda+\omega, & \varepsilon_7 &= -2\lambda+3\omega, & \varepsilon_8 &= 2\lambda+3\omega,
	\end{align}
	and the corresponding eigenvectors are
	\begin{subequations}
		\begin{align}
			\ket{\varepsilon_0} &= (1,0,0,0,0,0,0,0,0)^\top, \\
			\ket{\varepsilon_1} &= (0,0,0,0,0,-1,0,1,0)^\top, \\
			\ket{\varepsilon_2} &= (0,-1,0,1,0,0,0,0,0)^\top, \\
			\ket{\varepsilon_3} &= (0,0,0,0,0,0,0,0,1)^\top, \\
			\ket{\varepsilon_4} &= (0,1,0,1,0,0,0,0,0)^\top, \\
			\ket{\varepsilon_5} &= (0,0,0,0,0,1,0,1,0)^\top, \\
			\ket{\varepsilon_6} &= \left(0,0,f_1(\lambda,\omega),0,g_1(\lambda,\omega),0,1,0,0\right)^\top, \\
			\ket{\varepsilon_7} &= \left(0,0,f_2(\lambda,\omega),0,g_2(\lambda,\omega),0,1,0,0\right)^\top, \\
			\ket{\varepsilon_8} &= \left(0,0,f_3(\lambda,\omega),0,g_3(\lambda,\omega),0,1,0,0\right)^\top,
		\end{align} 
	\end{subequations}
	where
	\begin{align}
		f_\ell(\lambda,\omega) &= -1+\frac{2}{\lambda^2} \left(\omega-\frac{1}{2}\,{\rm Root[q,\ell]}\right)^2, \\
		g_\ell(\lambda,\omega) &= \frac{1}{\lambda} \left(-\sqrt{2}\, \omega + \frac{1}{\sqrt{2}}\, {\rm Root} [q,\ell]\right),
	\end{align}
	and ${\rm Root}[q, \ell]$ denotes the $\ell$-th root of the parameterized polynomial function
	\begin{equation}
		q(x;\lambda,\omega) \coloneq 8\omega(\lambda^2-\omega^2) + (-4\lambda^2+12\omega^2) x - 6\omega x^2 + x^3.
	\end{equation}
	Then, it is straightforward to calculate the Gibbs state of $\hat{H}_{\rm CHO}$ by substituting these eigenstates and their corresponding eigenenergies into \erf{rho:th:exact:gen}.

	\bibliography{references}
	
\end{document}